\documentclass{article}

\usepackage{PRIMEarxiv}
\usepackage{ulem}
\usepackage{amsmath}
\usepackage{bigints}
\usepackage[utf8]{inputenc} 
\usepackage[T1]{fontenc}    
\usepackage{hyperref}       
\usepackage{url}            
\usepackage{booktabs}       
\usepackage{amsfonts}       
\usepackage{nicefrac}       
\usepackage{microtype}      
\usepackage{lipsum}
\usepackage{tabularx}
\usepackage{fancyhdr}       
\usepackage{xcolor}
\usepackage{graphicx}       
\graphicspath{{media/}}     
\newcommand{\boldx}{\boldsymbol{x}}

\title{Chemotaxis-inspired PDE models of airborne infectious disease transmission: epidemiologically-motivated mathematical and numerical analyses

\author{
  Alex Viguerie \\
  Dipartimento di Scienze Pure e Applicate \\
  Universit\`a degli Studi di Urbino Carlo Bo \\
  Via Sant'Andrea 34 \\
  Urbino, PU 61029, Italy\\
  \texttt{alexander.viguerie@uniurb.it} \\
\and 
  \textbf{Malú Grave} \\
  Dept. of Civil Engineering\\
 Fluminense Federal University \\
  Rua Passo da Pátria, 156 \\ RJ 24210-240, Niterói, Brazil \\
  \texttt{malugrave@id.uff.br}  \\
  \and
 \textbf{Alvaro L.G.A. Coutinho} \\
  Dept. of Civil Engineering\\
  COPPE/Federal University of Rio de Janeiro \\
  P.O. Box 68506, RJ 21945-970 \\
  Rio de Janeiro, Brazil \\
  \texttt{alvaro@nacad.ufrj.br} \\
\and
  \textbf{Alessandro Veneziani} \\
Department of Mathematics \\ Department of Computer Sciences\\
Emory University \\
 200 Dowman Drive \\
Atlanta, GA (USA)\\
  \texttt{ale@mathcs.emory.edu} \\
  \and 
  \textbf{Thomas J.R. Hughes} \\
Oden Institute for Computational Engineering and Sciences \\ The University of Texas at Austin \\
 201 E. 24th Street \\
Austin, TX (USA)\\
}
}
  
\begin{document}
\maketitle

\begin{abstract}
Partial differential equation (PDE) models for infectious diseases, while less common than their ordinary differential equation (ODE) counterparts, have found successful applications for many years. Such models are typically of reaction-diffusion type, and model spatial propagation as a diffusive process. However, given the complex nature of human mobility, such models are limited in their ability to describe airborne infectious diseases in human populations. Recent work has advocated for the inclusion of an additional chemotaxis-type term as an alternative; spatial propagation of infection fronts is assumed additionally to flow from low-to-high concentrations of susceptible populations. The present work extends the study of such models by providing an epidemiologically interpretable analysis, directly connecting model behavior to information readily available to policymakers. In particular, we derive a spatially-aware basic reproduction number, which accounts for spatial heterogeneity in population density. Furthermore, we discuss several important aspects concerning the numerical solution of the model, including the introduction of a stabilization scheme. Finally, we perform a series of simulation studies in the Italian region of Lombardy (severely affected by the COVID-19 outbreak in 2020) and in the US state of Georgia, in which we demonstrate the model's potential to better capture important spatiotemporal dynamics observed in real-world data compared to pure reaction-diffusion models. 
\end{abstract}


\section{Introduction}
Over the past few decades, mathematical tools and models in public health and epidemiology have become increasingly prevalent. Beyond the obvious role of forecasting the progression of infectious diseases, these models have myriad other applications, such as evaluating intervention strategies in terms of costs and benefits, analyzing diverse scenarios to inform decision-makers, and guiding resource allocation \cite{knight2016bridging, fischer2016cdc}. While the COVID-19 pandemic has certainly renewed interest in this domain, it is important to note that mathematical models have been extensively used in epidemiology and public health for many years \cite{knight2016bridging, fischer2016cdc}, yielding numerous notable successes \cite{njeuhmeli2019using}. For more recent developments in epidemiological modeling and related computational aspects, readers may refer to \cite{kuhlcomp, zohdi2022}.

\par Of all the mathematical approaches for modeling infectious disease, \textit{compartmental models} based on systems of ordinary differential equations (ODEs) are the most commonly employed. This preference owes to a widely used special case of the seminal Kermack–McKendrick model \cite{kermack1927contribution} (see \cite{breda2012formulation} for a modern treatment), in which the population is divided into compartments according to disease stages. Such models are straightforward to implement, relatively easy to analyze, and have low computational overhead. However, they do not naturally incorporate spatial dynamics.

\par To this end, extensions to the compartmental framework based on partial differential equations (PDEs) have been proposed \cite{murray2003mathematical, schiesser2018mathematical}. In many cases, these PDE-based models treat the spatial spread of a disease as a Fickian diffusive process, meaning the disease propagates along the negative gradient of infection density (i.e., from areas of high infection density to areas of low infection density). PDE models are particularly well-suited for modeling vector-borne diseases and diseases affecting plants or wildlife \cite{murray2003mathematical, keller2013numerical, miranda2021scaling, bai2018reaction}. Reaction–diffusion models have also been used to study infectious diseases in human populations. Although a considerable amount of recent work in this area focuses on COVID-19 \cite{viguerie2020diffusion, viguerie2021simulating, viguerie2022coupled, grave2022modeling, bertrand2021least, bertaglia2021hyperbolic, guglielmi2022delay, auricchio2022well}, PDE models have similarly been applied to other diseases or discussed in more general contexts \cite{yamazaki2018threshold, li2009modeling, albi2022kinetic, ramaswamy2021comprehensive}.

\par Despite these efforts, the use of PDEs to model infectious diseases in human populations remains limited for several reasons. Compared to ODEs, the numerical solution of PDEs is substantially more involved, requiring considerable computational resources, careful mesh generation, spatial discretization schemes, and effective preconditioners for solving potentially large linear systems at each time-step. Furthermore, the mathematical analysis of PDEs can be intricate, and, crucially, arcane to public health practitioners. Practical application of mathematical models in epidemiology often involves deriving interpretable quantities, such as the basic and control reproduction numbers, which can then be used by practitioners to inform decision-making. While standard methods and derivations exist for ODEs, even the \textit{definition} of analogous quantities is not straightforward for PDEs \cite{auricchio2022well,albi2022kinetic}.

\par From a practical standpoint, human mobility is complex and occurs over multiple scales \cite{bhouri2021covid, findlater2018human, belik2011natural, tizzoni2014use, wesolowski2016connecting}. These dynamics significantly influence the spatial spread of infectious diseases and may not be captured adequately by simple diffusion. Another challenge involves incorporating nonlocal dynamics, which are difficult to represent through diffusion alone. A major advantage of ODE models is their ability to introduce additional stratifications that couple distinct geographic areas; accounting for long-range transmission thus becomes straightforward. Ironically, then, despite lacking an explicit spatial component, ODE models can sometimes offer a more tractable spatiotemporal perspective on disease spread \cite{linka2020safe, bhouri2021covid, miranda2021scaling, gatto2020spread, sattenspiel1995structured, balcan2009multiscale}.

\par Yet, using ODE-based stratifications to capture spatial effects has inherent limitations. Most obviously, one cannot achieve a continuous representation of spatial dynamics with this approach. Moreover, the structure of the spatial stratification is not always obvious. A common practice is to divide regions along political boundaries, such as states within a country \cite{gatto2020spread, linka2020safe}, but these boundaries may not reflect the most relevant descriptors of population distribution. Substantial heterogeneity often exists within such large areas, and increasing spatial resolution to account for this heterogeneity necessarily requires introducing additional model stratifications. As the number of stratifications grows, the number of parameters needed to define interregional interactions increases quadratically, and obtaining suitable data to calibrate them becomes difficult or impossible. Consequently, this approach quickly becomes infeasible if one seeks high spatial resolution.

\par PDE models, for all their challenges, have the advantage of providing richer spatial information than ODE models while requiring fewer parameters - an important motivation for their further development. To capture nonlocal transmission processes, extensions of reaction–diffusion models incorporating bilaplacian \cite{murray2003mathematical} or fractional diffusion \cite{gustafson2017fractional, zhao2023spatiotemporal, el2020stochastic, hoan2020new} have been proposed. Other variants include models with additional advection terms (possibly defined on a network structure \cite{bertaglia2021hyperbolic, albi2022kinetic, ramaswamy2021comprehensive}) or mass-transfer operators \cite{grave2022modeling}. 

\par Notwithstanding these extensions, it is unclear whether an airborne infectious disease in a human population should truly \textit{diffuse}, i.e., spread from high-infection-density regions to low-infection-density regions. Data from COVID-19 suggests a strong correlation between population density and transmission, with higher transmission clustering in dense population centers \cite{sy2021population}. Spatiotemporal transmission patterns appear to be influenced by factors such as highway networks \cite{souch2021interstates} and population density \cite{hu2013scaling, d2024spatial}, factors not naturally captured by diffusion. Thus, the pure diffusion model might be incomplete at best, even for describing local transmission dynamics. 

\par Recent work indicates that introducing a term inspired by \textit{chemotaxis} could more accurately describe the spatial–temporal propagation of infectious diseases \cite{colli2024chemotaxis}. This term posits that an infectious disease should move along the \textit{positive gradient} of susceptible density—i.e., from regions of lower susceptible population to those with higher susceptible population. Mathematically, this setup resembles chemotaxis models in which a biological species migrates in response to attraction or repulsion triggered by a chemical stimulus \cite{gajewski1998global, bellomo2022chemotaxis, marinoschi2013well, negreanu2014two, chalub2004kinetic, kirk2009parallel}. In \cite{colli2024chemotaxis}, a chemotaxis-like model for airborne infectious disease was introduced and shown to be well-posed. 

\par The present article extends that analysis in three major ways. First, we aim to provide an interpretation that connects the model’s physical behavior to its inputs in an epidemiologically meaningful manner, which should be useful for those focused on applications. In particular, we aim to define an extended reproduction number $R_0$ that accounts for the effects of spatial population heterogeneity on transmission. Second, we expand upon the preliminary simulations in \cite{colli2024chemotaxis} by considering both longer time horizons and additional regions of interest. Third, we offer further discussion on practical considerations for numerically solving the model, whose increased complexity naturally introduces additional computational challenges.

\par The article is organized as follows. We begin by recalling the mathematical model, its fundamental properties and underlying rationale. After establishing essential notation and formality, we conduct a mathematical analysis that elucidates how the model’s behavior relates to standard epidemiological concepts. In particular, we derive analogs of the well-known reproduction number, incorporating the effects of spatial heterogeneity—potentially useful for identifying vulnerable areas in intervention planning. We then offer a brief discussion of key aspects regarding the numerical solution of the model. Next, we present two simulation studies: one in the Italian region of Lombardy and another in the U.S. state of Georgia. By comparing the results qualitatively with COVID-19 surveillance data, we show that the chemotaxis model delivers a more accurate description of spatiotemporal disease spread than a pure diffusion model, even with time-independent parameters. We conclude with a summary of the main findings and propose several avenues for future research in this area.

\section{Mathematical Model}
We introduce the model first introduced in \cite{colli2024chemotaxis}. While we will recall some aspects of the model phenomenology, we refer the reader to \cite{colli2024chemotaxis} for the full derivation and explanation. 

\par Let \(\Omega \subset \mathbb{R}^2\) be a simply connected domain, and let \(i(x,t) \in H^2(\Omega)\) denote the \textit{density} of infected individuals at a point \(x\) in \(\Omega\) at time \(t\).

Following the derivation shown in \cite{colli2024chemotaxis}, one arrives at the following partial differential equation for  $i(x,t)$:
\begin{equation}
\partial_t i(x,t) \;=\; \Lambda(x,t) \;-\; \phi\,i(x,t) \;-\; \nabla \cdot \mathbf{j}
\end{equation}
throughout \(\Omega\) for all \(t\). $\Lambda(x,t)$ denotes the new infections, or \textit{incidence} at $(x,t)$, and $\mathbf{j}$ a flux term; both terms must be specified. For the former, we employ a standard frequency-dependent formulation:
\begin{equation}
\Lambda(x,t) \;=\; \frac{\beta\, i}{N}\, s,
\end{equation}
where \(s\) is the susceptible population at \((x,t)\), \(\beta\) is the \textit{contact rate} (with units 1/Time), and \(N\) is the total living population at \((x,t)\). We note that other choices for \(\Lambda\) may be preferable in certain contexts, including density-dependent or Holling-type formulations, among others. We refer the reader to \cite{blackwood2018introduction} for more details. We will next focus on the selection of \(\mathbf{j}\).

\subsection{Flux definition}

 We consider that the spatial evolution of the infectious disease also depends on two factors: a diffusion component arising from a Brownian motion (see \cite{salsa2022partial, colli2024chemotaxis}) as well as more complex processes, not captured by a simple random walk.

\par We assume the following:
\begin{itemize}
    \item The spatial propagation of an infectious disease depends on human mobility, in particular how this mobility relates to \textit{contacts}. In general, contacts will increase moving from regions of lower to higher population density.
    \item Higher levels of infection should, in general, serve to \textit{increase}, not decrease, the spatial propagation of the disease \textit{provided sufficient availability of susceptible individuals.}
\end{itemize}

Combining these points, we therefore postulate that \textit{an airborne infectious disease should propagate according to the density of susceptible individuals}, and in particular \textit{will generally move from regions with a lower concentration of susceptible individuals to a higher concentration of susceptible individuals}, with the rate of this movement also governed in part by the concentration of infected individuals in the area. These general principles are supported in various empirical studies (see e.g. \cite{sy2021population, wong2020spreading, souch2021interstates, hu2013scaling, d2024spatial}). This motivates the following definition for $\mathbf{j}$:

\begin{equation}\label{altFlux}
\mathbf{j} = -\nu_i \nabla i + \mu_i \chi(i) \nabla s, 
\quad \chi(i) \equiv \dfrac{i}{1 + i/C_0}
\end{equation}
where $C_0 > 0$ (see \cite{colli2024chemotaxis}). The parameter $C_0$ functions as a saturation capacity term. Observe:
\begin{equation}\label{functionExplanation}
    \chi(i) \approx i,\,\, i<<C_0,\, \text{and}\,\,\lim_{i\to \infty } \chi(i) = C_0.
\end{equation}

The full model then reads as follows:
\begin{alignat}{2}
   \label{eqn2S}
\partial_t s &=-  \frac{\beta i}{s+i+r}s  \qquad\qquad\qquad\qquad &&\text{ in } \Omega \times \lbrack 0,\,T),\\
   \label{eqn2I}
   \partial_t i &=  \frac{\beta i}{s+i+r}s -\phi_i i + \nabla \cdot \left( \nu_i \nabla i -\mu_i \chi(i) \nabla s\right)   &&\text{ in } \Omega \times \lbrack 0,\,T), \\
   \label{eqn2R}
   \partial_t r &=  \phi_i i  &&\text{ in } \Omega \times \lbrack 0,\,T), \\
   \label{eqn2BC}
\partial_{\boldsymbol{n}}s&=0,\,\partial_{\boldsymbol{n}}i = 0,\,\partial_{\boldsymbol{n}}r = 0 &&\text{ on } \partial \Omega \,\, \forall t,\\
   \label{eqn2IC}
s(x,0)&=s_0(x),\,i(x,0)=i_0(x),\,r(x,0)=r_0(x) &&\text{ at } t=0.
\end{alignat}

\par  We  do not consider diffusion (or chemotaxis) in the $s$ and $r$ compartments, as we do not feel it is realistic to model population movement in these compartments as tending towards a spatial equilibrium. Furthermore, human mobility over the time scales considered is largely transient—people will regularly return to a home location \cite{murray2003mathematical, schneider2013unravelling}. Therefore, it is more accurate to interpret the model as describing the \textit{contacts} resulting from mobility, rather than mobility itself. This reasoning is further elaborated in \cite{colli2024chemotaxis}.

\begin{table}
\begin{center}
\begin{tabular}{ |c|c|c| } 
\hline
Parameter &  Name  &  Unit  \\
\hline\hline
$\beta$ & Contact rate & Days$^{-1}$ \\ \hline
$\phi$ & Removal rate & Days$^{-1}$ \\ \hline
$\nu_j$ & Diffusion coefficient, compartment $j$ & km$^2$ $\cdot$ Days$^{-1}$  \\ \hline
$\mu_j$ & Chemotaxis coefficient, compartment $j$ & km$^2$ $\cdot$ Persons $\cdot$ Days$^{-1}$  \\ \hline
$C_0$ & Saturation/capacity term &  Persons $\cdot$ km$^{-2}$  \\
\hline
\end{tabular}
\caption{Description of parameters of the chemotaxis model \eqref{eqn2S}-\eqref{eqn2IC}.}
\label{tab:ChemotaxisModelParameters}
\end{center}
\end{table}

\section{Mathematical analysis}
In this section, we present some mathematical results regarding the system \eqref{eqn2S}-\eqref{eqn2IC}.

\subsection{Variational formulation}
Denote
$V:=L^2 (0,T; H^1(\Omega))$. Further, denote the $L^2$ inner product over $\Omega$ as:

\begin{equation}\label{L2Product}
    (u,v) = \int_{\Omega} u\, v\, d\boldx,\\
\end{equation}
\par Multiplying the system \eqref{eqn2S}-\eqref{eqn2IC} by $(w,v,l)\in V \times V\times V$ and integrating gives:

\begin{align}\begin{split}\label{varForm}
 (\partial_t s,v) &= -\left(\dfrac{\beta i}{s+i+r}s,v\right)  \\
 (\partial_t i,w) &= \left(\dfrac{\beta i}{s+i+r}s,w\right) - (\phi_i i,w) + \left(\nabla\cdot \left(\nu_i \nabla i  - \mu_i \chi(i) \nabla s\right),w\right) \\
 (\partial_t r,l) &=  (\phi_i i,l).
 \end{split}\end{align} 
 
 \par Integrating by parts yields the following variational problem:
 \par \textbf{Problem 1:} \textit{Find $(s,i,r)\in V \times V \times V$ such that for all $(v,w,l) \in V \times V \times V$:}
\begin{align}\label{varFormS}
( \partial_t s,v) &= -\left(\dfrac{\beta i}{s+i+r}s,v\right)  \\
 \label{varFormI}
 (  \partial_t i,w) &= \left(\dfrac{\beta i}{s+i+r}s,w\right) - (\phi_i i,w) - \left(\nu_i \nabla i,\nabla w \right)  + \left(\mu_i \chi(i) \nabla s,\nabla w\right)   \\
 \label{varFormR}
  ( \partial_t r,l) &=  (\phi_i i,l).
\end{align} 
Note the boundary terms resulting from the integration by parts vanish from the homogeneous Neumann boundary conditions.

\subsection{Mathematical results}

\subsubsection{Relevant Results}

Here we restate a series of results from \cite{colli2024chemotaxis}, which will be used throughout the present.

\begin{itemize}
\item $(\nu_s, \nu_i, \nu_r) \in (L^\infty(Q))^3, 0 \leq \nu_m \leq \nu_s, \nu_i, \nu_r \leq \nu_M$ a.e. in $Q$ (where $\nu_M$ is a constant).

\item $\beta \in L^\infty (Q), \ 0 \leq \beta \leq \beta_M$ a.e. in $Q$
(where $\beta_M$ is a constant).

\item $\phi_i >0$ is a constant and \footnote{The paper \cite{colli2024chemotaxis} is more general, but we stick with this reasonable choice for this function.}
 $\chi(i) = \dfrac{i}{1+i/C_0}$.

\item The initial conditions are such that:

$(s_0, i_0, r_0) \in (L^2(Q))^3, 0 < s_m \leq  s_0 \leq s_M$ a.e. in $Q$ (where $s_m, s_M$ are constant), $i_0, r_0 \geq 0$.

\end{itemize}

\par \textbf{Theorem P1} Under the assumptions listed above, there exists a triplet:

$$
(s,i,r) \in H^1(H^{1\prime}) \cap L^2(H^1)
$$

weak solution of our problem, such that a.e. in $Q$

$$
s_m \leq s \leq s_M, i \geq 0, r \geq 0
$$

\noindent where $\dfrac{si}{s+i+r}$ is intended to be $=0$ whenever $s+i+r=0$.

Moreover, for $N = s + i + r$ and $N_0 = s_0 + i_0 + r_0$, then
we have the conservation property:

$$
\displaystyle{\int\limits_\Omega} N dx = \displaystyle{\int\limits_\Omega} N_0 dx. 
$$

\par \textbf{Theorem P2}

Under the same assumptions of Theorem P1, and in addition:

\begin{itemize}
    \item $\nu_s$ constant
    \item $\beta$ in $H^1(\Omega)$ or $L^2(0,T;H^1(\Omega))$ if time dependent
    \item $s_0 \in H^2(\Omega)$ with $\nabla s \cdot \mathbf{n} = 0$ on $\partial \Omega$
\end{itemize}
then (a) the weak solution of Theorem P1 is unique and (b) $s$ is a strong solution to the problem, such that
$$
\| s \|_{H^1(0,T;H^1)\cap L^\infty(0,T,H^2)} \leq C.
$$

\par \textbf{Remark}
Notice that an immediate consequence of the Theorems above states that for every $t>0$
$$
\int_\Omega i d \boldx \leq \int_\Omega N_0 d \boldx - A(\Omega) s_m
$$
where $A(\Omega)$ is the area of the region $\Omega$.

\par \textbf{Theorem P3}

Under the assumptions of Theorem P2 and $r_0 \in H^1(\Omega)$, the solution of the previous theorem 
converges for $\nu_s, \nu_r \rightarrow 0$ in the appropriate topologies  
to a triplet $(s,i,r)$ that we identify as the weak solution of our problem for $\nu_r = \nu_s = 0$.

\subsubsection{Analysis of stability and growth of the population.} 

In this Section, we refer to the case $\nu_s = \nu_r = 0$ under the assumptions of Theorem P3.
We provide a series of results related to the behavior of total population $N=s+i+r$ in the $L^2$ norm.
First of all, note that from the previous results, we promptly have the inequality:
$$
\int_\Omega N_0 \boldx \equiv \| s + i + r \|_{L^1(\Omega)} \leq A(\Omega) \| s + i + r \|_{L^2(\Omega)} 
$$

Adding the equations \eqref{varFormS}-\eqref{varFormR}, we obtain:

\begin{align}\label{testFnl}
  (\partial_t s,v)+ (\partial_t i,w)+(\partial_t r,l ) &= \left(\dfrac{\beta i}{s+i+r}s,v-w\right) +\phi_i (i,l-w) - \nu_i(\nabla i,\nabla w) + \mu_i (\chi(i)\nabla s, \nabla w).
\end{align}
Letting $v=w=l=s+i+r$,
\begin{align}\label{testFnl_1}
\dfrac{1}{2}\partial_t \| s+i+r\|_{L^2}^2 &= \left(\mu_i \chi(i)\nabla s - \nu_i \nabla i,\,\nabla\left(s+i+r\right)\right) \\
\label{testFnl2}
&= \bigintss_{\Omega} \begin{pmatrix} \nabla s \\ \nabla i\\ \nabla r\end{pmatrix}^T 
\begin{pmatrix}
\mu_i \chi(i)I & -\nu_iI & 0 \\
\mu_i \chi(i)I & -\nu_iI & 0 \\
\mu_i \chi(i)I & -\nu_iI & 0 \\
\end{pmatrix} \begin{pmatrix} \nabla s \\ \nabla i\\ \nabla r\end{pmatrix}  d\boldx .\end{align}

Letting:
\begin{equation}
A=\begin{pmatrix}
\mu_i \chi(i)I & -\nu_iI & 0 \\
\mu_i \chi(i)I & -\nu_iI & 0 \\
\mu_i \chi(i)I & -\nu_iI & 0 \\
\end{pmatrix}, \qquad \boldsymbol{u} = \begin{pmatrix} \nabla s \\ \nabla i \\ \nabla r \end{pmatrix}, \end{equation}
we can write \eqref{testFnl2} as:
\begin{equation}\label{testFnl3}
\dfrac{1}{2}\partial_t \| s+i+r\|_{L^2}^2 = \int_{\Omega} \boldsymbol{u}^{T} A \boldsymbol{u} \, d\boldx =
\int_{\Omega} \boldsymbol{u}^{T} A_{symm} \boldsymbol{u} \, d\boldx.
\end{equation}
where $A_{symm} \equiv (A+A^T)/2$.
By direct inspection:
\eqref{testFnl_1}: \begin{align}\label{quadFormSymPart}
\dfrac{A+A^T}{2}=
\begin{pmatrix}
 \mu_i \chi(i)I & ( \mu_i \chi(i)I - \nu_i I)/2  & \mu_i \chi(i) I/2 \\
( \mu_i \chi(i) I - \nu_i I)/2 & -\nu_iI & -\nu_i I/2 \\
\mu_i \chi(i) I/2 & -\nu_i I/2 & 0 \\
\end{pmatrix}
\end{align}
\par Explicit computations show that the eigenvalues of \eqref{quadFormSymPart} are given by:
\begin{equation}
\lambda_0 = 0,\, \lambda_{\pm} = \frac{\mu_i \chi(i) - \nu_i }{2} \pm \frac{\sqrt{3 \left(\mu_i^2 \chi^2(i) + \nu_i^2 \right)}}{2} = \dfrac{\nu_i}{2}\left\lbrack \left(R_s-1\right) \pm \sqrt{3}\sqrt{R_s^2+1}\right\rbrack,
\end{equation}
where:
\begin{equation}\label{RsDefn}
    R_s=\dfrac{\mu_i \chi(i)}{\nu_i}\geq0,
\end{equation}
with the nonnegativity following from the nonnegativity of $\mu_i$, $i$, and $\nu_i$ respectively. Observe that:
\begin{equation}
\lambda_- \lambda_+ = \frac{\nu_i^2}{4} \left\lbrack (R_s-1)^2 - 3\left(R_s^2+1\right) \right\rbrack = -\frac{\nu_i^2}{2}\left(R_s^2+R_s+1\right) < 0,
\end{equation}
which, from the nonnegativity of $R_s$, implies that the two nonzero eigenvalues of $(A+A^T)/2$ always have opposite sign. Furthermore, the nonnegativity of $R_s$ also guarantees the elementary inequalities:

\begin{equation}
    \sqrt{R_s^2+1} \leq \sqrt{R_s^2+2R_s +1} = \sqrt{(R_s+1)^2}=R_s+1
\end{equation}
and
\begin{equation}
    \sqrt{R_s^2+1} \geq \sqrt{R_s^2} = R_s.
\end{equation}
These can be used to show that:
\begin{equation}
\frac{\nu_i}{2}\left\lbrack (1-\sqrt{3})R_s - (1+\sqrt{3})\right\rbrack \leq \lambda_- \leq \frac{\nu_i}{2} \left\lbrack (1-\sqrt{3})R_s -1 \right\rbrack \leq -\frac{\nu_i}{2}
\end{equation}
and
\begin{equation}
\frac{\nu_i \left(\sqrt{3}-1\right)}{2} \leq \lambda_+ \leq \frac{\nu_i}{2} \left\lbrack (1+\sqrt{3})R_s - (1-\sqrt{3}) \right\rbrack.
\end{equation}
Moreover, the following facts can be proven:
\begin{itemize}
\item $|\lambda_+| = |\lambda_-| $ if and only if $R_s=1$;
\item $|\lambda_+| > |\lambda_-| $ if and only if $R_s>$1; 
\item $|\lambda_+| < |\lambda_-| $ if and only if $R_s<$1.
\end{itemize}

Clearly, if $R_s > 1$ ($R_s < 1$) everywhere, then $\| s+i+r\|_{L^2}^2$ exhibits growth (decay). 
According to this definition $R_s$ depends on space through $i$, so it can hardly be applied throughout the region.
However, $0 \leq \chi(i) \leq C_0$, so that $R_s \leq R_s^* \equiv \mu_i C_0 /\nu_i$.
Consequently, if $R_s^* < 1$, we expect a global decay.

This analysis establishes that, in regimes where chemotactic drift dominates over diffusion, $\| N \|_{L^2}$ can grow despite mass conservation in $\| N \|_{L^1}$, implying spatial aggregation (and possible instability arising from sharp gradients). Note that as this result concerns the entire population $N$, and not the infected population specifically, its epidemiological interpretation is not immediate. However, it provides a Peclet-type indicator for characterizing model behavior.


\subsection{Analysis of the time evolution of the infection population.}

Denote:

$$
\alpha(t) \equiv \int_0^t \dfrac{2 \mu_i^2 C_0^2 }{\nu_i} \| \nabla s \|_0^2 dt +  \| i_0 \|_0^2
$$

\noindent \textbf{Theorem 2A}  
Under the assumptions of Theorems P1, P2, (P3) and assuming $\varphi_i$ constant, we have for any $t>0$
\begin{equation}\label{eq:ven1}
\left\{ 
\begin{array}{lll}
\| i\|_0^2 (t) \leq \alpha(t) & \mathrm{for} & \varphi_i \geq \beta \\
\| i\|_0^2 (t) \leq \alpha(t) + (\beta - \varphi_i) \int_0^t \alpha(\tau) e^{(\beta - \varphi_i)(t-\tau)} d \tau & \mathrm{for} & \beta > \varphi_i 
\end{array} \right.
\end{equation}

\textbf{Proof.} 

Let us consider  the equation for the infected in \eqref{varFormI} with $w=i$ as a test function.
Recall that $s/(s+i+r) \leq 1$ for the positivity of the components of the triplet and $\chi(i) \leq C_0$. We obtain:

\begin{equation}\label{eq:VenProofFirstEq}
\frac{1}{2} \partial_t \|i\|_{L^2}^2 
\leq \left(\beta - \phi_i\right)\|i\|_{L^2}^2 -\nu_i \| \nabla i\|_{L^2}^2 + \mu_i (\chi(i) \nabla s,\nabla i ) \\
\leq \left(\beta - \phi_i\right)\|i\|_{L^2}^2 -\nu_i \| \nabla i\|_{L^2}^2 + \mu_i C_0 \| \nabla i\| \|\nabla s \|, 
\end{equation}

Noting that
$$
\mu_i C_0 \| \nabla i\| \|\nabla s \| \leq \mu_i C_0 \epsilon \| \nabla i\|^2 + \dfrac{\mu_i C_0 }{4 \epsilon} \|\nabla s \|^2
$$
and selecting  $\epsilon$ such that $\epsilon = \nu_i/(C_0 \mu_i)$, we obtain

\begin{equation}\label{eq:VenProofFirstEq2}
\partial_t \|i\|_{L^2}^2 
\leq 2\left(\beta - \phi_i\right)\|i\|_{L^2}^2 + \dfrac{2 C_0^2 \mu_i^2}{\nu_i} \|\nabla s \|^2, 
\end{equation}

If $\beta - \phi_i > 0$, we obtain the thesis by applying the Gronwall Lemma. If $\beta - \phi_i \leq  0$ we simply drop 
the first term in the inequality and integrate in time.

QED

This Theorem interestingly correlates the growth of the infected to the space gradient of the susceptibles in the presence of chemotactic 
diffusion. Even if the bounds are not sharp, local growth of the infected may be reduced if the distribution of susceptibles is quite uniform. In such a case, the local behavior resembles that of a pure ODE system at each point. 

In this next Theorem, we take advantage of the results of Theorem P2 and the fact that $s$ is a strong solution. In fact, we assume that 
\begin{equation}\label{assumptionRegularity1}
    \Delta s > - \dfrac{\phi_i}{\mu_i}.
\end{equation}
The assumption states that the distribution of the susceptible population in space is sufficiently smooth. While we cannot predict \textit{a priori}  that this condition is true, we can verify that it holds \textit{a posteriori}. Furthermore, if $s_0 >> i_0$ (a standard assumption in many applications, e.g., computing $R_0$ from early case count data), one may use $\Delta N_0 $, readily available, as a surrogate in practice.

Define:
\begin{equation}    
R_0 = \dfrac{  \beta }{\phi_i + \dfrac{\mu_i \Delta s}{2} }.
\end{equation}

\noindent \textbf{Theorem 2B}  
Let us postulate  the same assumptions of Theorem 2A and that \eqref{assumptionRegularity1} holds.
Let us assume also that $i$ is small enough ($i \ll C_0$)  to approximate $\chi(i) \approx i$. Then,
the infected population $i$ is guaranteed to decrease in the $L^2$ sense if $R_{0}<1$ over $\Omega$.

\textbf{Proof.} 

The assumption on the smallness of $i$ clearly applies at the beginning of the outbreak, when most of the population is 
represented by $s$. This is exactly when it is critical to find predictive indexes of the growth.
Proceeding as in the previous Theorem,
notice that
$$
\mu_i (\chi(i) \nabla s,\nabla i ) \approx \mu_i (i \nabla s, \nabla i) = \frac{\mu_i}{2} (\nabla s, \nabla i^2) =    -\frac{\mu_i}{2} \int_\Omega i^2 \Delta s,  
$$
with boundary terms disappearing due to the homogenous Neumann conditions. Then, we obtain the inequality:
$$
\frac{1}{2} \partial_t \|i\|_{L^2}^2 
\leq \left(\beta - \phi_i - \frac{\mu_i \Delta s}{2} \right)\|i\|_{L^2}^2. 
$$
The diffusive term weighted by $\nu_i$ was removed from the right hand side, being non-positive.
From here, we promptly obtain
$$
\frac{1}{2} \partial_t \|i\|_{L^2}^2 
\leq \left(R_0 -1\right)\left(\phi_i + \frac{\mu_i \Delta s}{2} \right)\|i\|_{L^2}^2. 
$$
as was to be shown.

\paragraph{On the Boundary Conditions}

We have assumed to work in an isolated region with homogeneous boundary conditions. 
If we set a portion of the boundary with (homogeneous) Dirichlet conditions, we do not have the theoretical results
of \cite{colli2024chemotaxis} in the background. However, we assume that similar results could be proven in that case (which is not clearly the scope of the present paper). Then, we have the Poincar\'e inequality
\begin{equation}\label{poincareIneq}
\| i \|_{L^2}^2 \leq C_{\Omega} \| \nabla i \|_{L^2}^2\, \, \, 
\end{equation}
at all $t$, for some constant $C_\Omega$. In practice, we may expect this to be the case if, for example, we consider a domain with at least one sufficiently remote, dispopulated region either within $\Omega$ or on $\partial \Omega$ such that we may fix $s=0,\,i=0,\,r=0$ in this region.

\par We can make the definition of $R_0$ with the following additional  condition on $s$:
\begin{equation}\label{assumptionRegularity}
    \Delta s > - {\mu_i} \left\lbrack \dfrac{\nu_i}{C_\Omega}  + \phi_i \right\rbrack,
\end{equation}
at all $(\boldx,t)$. Note that in general $\nu_i >> \mu_i$, so this  assumption is not overly stringent.
Define 
\begin{align}\label{spatialR0Defn}
R_{0,D} &= \dfrac{  \beta }{\phi_i + \dfrac{\mu_i \Delta s}{2} + \dfrac{\nu_i}{C_\Omega}  }.
\end{align}

\noindent \textbf{Theorem 2C.} Assuming \eqref{assumptionRegularity} holds, the infected population $i$ is guaranteed to decrease over a subdomain $\Omega_0$ in the $L^2$ sense if $R_{0,D}<1$ over $\Omega$.

\textbf{Proof.} 

The proof follows a similar strategy as in the previous Theorem, but in this case we do not drop the diffusive term, 
we just use the Poincar\'e inequality to bound it.

QED

\subsubsection{Local estimates}

\par While the above results hold over the entire domain, we may extend them locally. In the interest of epidemiological clarity, we assume additional regularity, allowing us to derive bounds in terms of spatially resolved quantities that can be readily interpreted. Namely, assume $\nabla i,\,\nabla s \in L^\infty (\Omega) $. As previously, we assume that $i << C_0$, such that we may consider $\chi(i) \approx i$. Consider a subdomain $\Omega_0 \subset \Omega$ and define for $\varepsilon>0$:

\begin{equation}
\Omega_{0,\varepsilon} 
\;:=\; 
\{\,x \in \Omega : \mathrm{dist}(x,\Omega_0) < \varepsilon\},    
\end{equation}
and let $\sigma:\Omega \to [0,1]$ be a smooth ``cutoff'' function satisfying
\begin{equation}\label{sigmaDefn}
  \sigma(x) 
  = 
  \begin{cases}
     1, & x \in \Omega_0,\\
     0, & x \notin \Omega_{0,\varepsilon},
  \end{cases}
  \qquad
  0 < \sigma(x) < 1 
  \;\;\text{only for }x \in \Omega_{0,\varepsilon}\!\setminus\!\Omega_0.
\end{equation}
Moreover, assume there is an integer $p \geq 1$ and a constant $C_\sigma>0$ so that
\begin{equation}\label{sigmaGrowth}
  \|\nabla \sigma\|_{L^\infty(\Omega)} 
  \;\le\;
  \frac{C_\sigma}{\varepsilon^{\,p}}.
\end{equation}

\textbf{Remark on $p$.} The exponent $p$ here encodes how rapidly $\sigma$ transitions from $1$ to $0$ over a strip of width $\varepsilon$. In many practical constructions, we choose $\sigma$ to vary linearly (or piecewise smoothly) from $1$ to $0$, and then $p=1$ because $\|\nabla \sigma\|\sim 1/\varepsilon$. More complicated schemes or domain geometries can alter this exponent, but $p=1$ is common in dimension $2$ or $3$ when $\sigma$ is simply a linear ramp-down.

\par The local argument follows a procedure analogous to the global version, but setting $w = \sigma^2 i$ in \eqref{varFormI}. Within $\Omega_0$, $\sigma=1$, yielding:

\begin{equation}
\dfrac{1}{2} \partial_t \| i \|_{L^2(\Omega_0)}^2 = (\beta - \phi_i) \| i \|_{L^2(\Omega_0)}^2  - \nu_i \| \nabla i  \|_{L^2(\Omega_0)}^2 + \dfrac{\mu_i}{2}(\nabla i^2,\nabla  s) _{\Omega_0}
\end{equation}
By integrating the last term above by parts and applying the Poincare and trace inequalities to absorb the boundary contributions, we obtain an estimate of the form:
\begin{equation}\label{locEst}
 \partial_t \| i \|_{L^2(\Omega_0)}^2 \leq \int_{\Omega_0} 2\left(\beta - \left(\phi_i + \dfrac{\mu_i \Delta s}{2} + \dfrac{\nu_i}{C_{\Omega_0}} + C_{\partial \Omega_0}\|\nabla s\|_\infty \right)\right)|i|^2 d\boldx .
\end{equation}
Over $\Omega_{0,\varepsilon}$, we also have contributions from the following terms:
\begin{equation}
\nu_i \int_{\Omega_{0,\varepsilon}} 2\sigma i \nabla i \cdot \nabla \sigma d\boldx, \,\,\text{and} \,\, \mu_i \int_{\Omega_{0,\varepsilon}} 2\sigma i^2 \nabla s \cdot \nabla \sigma d\boldx .
\end{equation}
From Cauchy-Schwarz:
\begin{equation}
 | \nabla i \cdot \nabla \sigma | \leq \| \nabla i \| \| \nabla \sigma \|,\,\,\,|\nabla s \cdot \nabla \sigma | \leq \| \nabla s \| \| \nabla \sigma \|,
\end{equation}
which, together with the ``Peter-Paul" inequality, yields:
\begin{equation}\label{peterPaul1}
    \big| \nu_i \int_{\Omega_{0,\varepsilon}} 2\sigma i \nabla i \cdot \nabla \sigma d\boldx \big| \leq \dfrac{\nu_i }{2\varepsilon} \int_{\Omega_{0,\varepsilon}} (2\sigma i)^2 d\boldx + \dfrac{\nu_i \varepsilon}{2} \int_{\Omega_{0,\varepsilon}} \|\nabla i\|^2 \| \nabla \sigma \|^2 d\boldx
\end{equation}
and
\begin{equation}\label{peterPaul2}
    \big| \mu_i \int_{\Omega_{0,\varepsilon}} 2\sigma i^2 \nabla s \cdot \nabla \sigma d\boldx \big| \leq \dfrac{\mu_i }{2\varepsilon} \int_{\Omega_{0,\varepsilon}} (2\sigma i)^2 d\boldx + \dfrac{\mu_i \varepsilon}{2} \int_{\Omega_{0,\varepsilon}} i^2\|\nabla s\|^2 \| \nabla \sigma \|^2 d\boldx.
\end{equation}
The first terms on the right-hand side can be easily bounded by:
\begin{equation}\label{stripContrib1}
\big|    \dfrac{1}{2\varepsilon} \int_{\Omega_{0,\varepsilon}} (2\sigma i)^2 d\boldx \big| \leq \dfrac{2 C_{\Omega_0}}{\varepsilon} \|\nabla i\|_{\infty}^2 \ell(\partial \Omega_0) \varepsilon = 2 C_{\Omega_0} \| \nabla i \|_{\infty}^2 \ell(\partial \Omega_0 ),
\end{equation}
where $\ell(\Omega_0)$ is the perimeter length of $\Omega_0$. For the remaining terms, we apply \eqref{sigmaGrowth} with $p=1$ and obtain:
\begin{equation}\label{stripContrib2}
\big|  \dfrac{\nu_i \varepsilon}{2} \int_{\Omega_{0,\varepsilon}} \|\nabla i\|^2 \| \nabla \sigma \|^2 d\boldx \big|  \leq  \dfrac{  \nu_i \varepsilon}{2} \|\nabla i \|_\infty^2  \dfrac{C_\sigma}{\varepsilon^2} \ell(\partial \Omega_0) \varepsilon = \dfrac{\nu_i C_\sigma}{2} \| \nabla i \|_\infty^2 \ell(\partial \Omega_0)
\end{equation}
and 
\begin{equation}\label{stripContrib3}
\big|  \dfrac{\mu_i \varepsilon}{2} \int_{\Omega_{0,\varepsilon}} i^2\|\nabla s\|^2 \| \nabla \sigma \|^2 d\boldx \big|  \leq  \dfrac{  \mu_i \varepsilon}{2} \| i \|_\infty^2 \|\nabla s\|_\infty^2 \dfrac{C_\sigma}{\varepsilon^2} \ell(\partial \Omega_0) \varepsilon = \dfrac{C_\sigma C_{\Omega_0} \mu_i }{2} \| \nabla i \|_\infty^2 \| \nabla s \|_\infty^2 \ell(\partial \Omega_0 ).
\end{equation}
By absorbing \eqref{stripContrib1}, \eqref{stripContrib2}, \eqref{stripContrib3} into \eqref{locEst}, we obtain a localized bound. We note that the local version of the bound, unsurprisingly, introduces additional dependencies on the local geometry and local function behavior; however, the qualitative interpretation of the bound remains unchanged.

\par \textbf{Remark 1.} The derived $R_0$ shows that, for the model \eqref{eqn2S}-\eqref{eqn2IC}, the behavior of the local basic reproduction number depends not only on the transmission and recovery terms, which are the same throughout $\Omega$, but additionally on $\nu_i$ and $\mu_i \Delta s$. Locally, $\nu_i$ serves to reduce $\|i\|_{L^2}$. 
\par The effect of directed spatial propagation is reflected in the factor $\mu_i \Delta s $. When $\Delta s > 0$, this term increases the denominator of $R_0$, reducing $\|i\|_{L^2}$; in contrast, when $\Delta s < 0$,  $R_0$ increases, reflecting the growth of $\|i\|_{L^2}$. As we assume a susceptible population, we may intuit $s$ as representing the overall population density of a given region. As such, $\Delta s >0$ indicates \textit{positive concavity} of the population density, implying the presence of a local \textit{minimum} in the vicinity. Similarly, $\Delta s <0 $ implies the opposite - the presence of a nearby local \textit{maximum}.  This is consistent with our basic intuition that infectious disease should propagate towards regions of increasing population density (and hence negative concavity). 
\par\textbf{Remark 2.} Note that a standard definition of $R_0$ for a model of this type is given by $\widetilde{R}_0 = \beta/\phi_i$. The expression \eqref{spatialR0Defn} shows that, even if $\widetilde{R}_0<1$ in all of $\Omega$, this may not be sufficient to guarantee that $\|i\|_{L^2}$ remains under control, as the spatial propagation of the infection is subject to additional effects arising from the population density patterns.
\par \textbf{Remark 3.} Note the $R_0$ \eqref{spatialR0Defn} only provides an upper bound on $\partial_t \| i\|_{L^2}^2$, and it is not necessarily the case that $R_0>1$ implies local growth of $\|i\|_{L^2}$. This is an important distinction from the $R_0$ defined for analogous ODE models.



\subsection{Numerical aspects}
We must discretize the problem \eqref{varForm} in both time and space. The subject of spatial and temporal discretization for PDEs is vast; herein, we will consider a Finite Difference (FD) discretization in time, combined with a Finite Element (FEM) discretization in space. We provide a brief overview of these methods in the context of \eqref{varForm} and refer the reader to, e.g., \cite{hughes2012finite, quarteroni2008numerical} for additional details.

\par \textbf{Time-discrete formulation and linearization.} 
Let $\overline{V}:= H^1(\Omega)$. Given the diffusion terms, we expect that an implicit time-marching scheme is necessary to ensure stability. For ease of explanation, we consider a Backward-Euler time discretization, and note that the following discussion \textit{holds} without loss of generality for more complex time-advancing schemes. Denoting $N = s + i + r$, the time-discrete version of \eqref{varForm} with time-step $\Delta t$ reads:

\noindent \textbf{Problem 2.} 
\textit{Find $(s_{t+1}, i_{t+1}, r_{t+1})$ in $\overline{V}\times\overline{V}\times\overline{V}$ such that for all $(v,w,l)$ in $\overline{V}\times\overline{V}\times\overline{V}$:} 
\begin{align}
\begin{split}\label{timeDisc}
\left( \frac{ s_{t+1}-s_t }{ \Delta t },v\right) 
&= 
-\beta \left(\frac{ s_{t+1} \,\widetilde{i} }{\widetilde{N}},v\right),\\
\left( \frac{ i_{t+1}-i_t }{ \Delta t },w\right) 
&= 
\beta \left(\frac{ i_{t+1}\,\widetilde{s} }{\widetilde{N}},w\right) 
- \phi_i (i_{t+1},w) 
- \nu_i (\nabla i_{t+1},\nabla w) 
+ \mu_i (\chi(i_{t+1})\,\nabla \widetilde{s},\nabla w),\\
\left( \frac{ r_{t+1}-r_t }{ \Delta t },l\right) 
&= 
\phi_i (i_{t+1},l).
\end{split}
\end{align}
Solving \eqref{timeDisc} requires that we linearize the $\widetilde{s},\,\widetilde{i},\,\widetilde{N}$ terms. A simple approach is the \textit{semi-implicit} formulation:

\noindent \textbf{Problem 3, semi-implicit time discretization:} 
\textit{Find $(s_{t+1}, i_{t+1}, r_{t+1})$ in $\overline{V}\times\overline{V}\times\overline{V}$ such that for all $(v,w,l)$ in $\overline{V}\times\overline{V}\times\overline{V}$:} 
\begin{align}
\begin{split}\label{timeDiscSemiImp}
\left( \frac{ s_{t+1}-s_t }{ \Delta t },v\right) 
&= 
-\beta \left(\frac{ s_{t+1}\,i_t }{N_t},v\right),\\
\left( \frac{ i_{t+1}-i_t }{ \Delta t },w\right) 
&= 
\beta \left(\frac{ i_{t+1}\,s_t }{N_t},w\right) 
- \phi_i (i_{t+1},w) 
- \nu_i (\nabla i_{t+1},\nabla w) 
+ \mu_i (\dfrac{i_{t+1}}{i_{t}+C_0}\,\nabla s_t ,\nabla w),\\
\left( \frac{ r_{t+1}-r_t }{ \Delta t },l\right) 
&= 
\phi_i (i_{t+1},l).
\end{split}
\end{align}
While semi-implicit schemes offer the advantage of requiring no nonlinear iterations, they are unsuitable for the current problem. To see why, let $v=w=l=1$ and add the three equations above to obtain:

\begin{align}\label{massImbalance}
\int_{\Omega} (N_{t+1} - N_t)\, d\boldx 
= 
\Delta t \,\beta \int_{\Omega} \frac{ i_{t+1}\,s_t \;-\; s_{t+1}\,i_t}{N_t}\, d\boldx.
\end{align}
Hence, this scheme does not respect the mass conservation established in Theorem P1. Assuming sufficient time regularity, we expand $s_{t+1},\,i_{t+1}$ in Taylor series:
\begin{equation}
s_{t+1} = s_t + \Delta t \,\partial_t s_t + \mathcal{O}(\Delta t^2), 
\quad 
i_{t+1} = i_t + \Delta t \,\partial_t i_t + \mathcal{O}(\Delta t^2).
\end{equation}
Substituting these expansions into \eqref{massImbalance} and taking absolute values:
\begin{align}\label{massImbalanceBound}
\| N_{t+1} - N_t \|_{L^1} 
= 
\Delta t\, \beta \left\| \Delta t \bigl( \partial_t i_{t}\, s_t - \partial_t s_t \, i_t \bigr) N_t^{-1} + \mathcal{O}(\Delta t^2) \right\|_{L^1} 
\leq 
C \,\Delta t^2.
\end{align}
Considering a fixed time interval $\lbrack 0,\,T\rbrack$ and letting $K = T/\Delta t$, from \eqref{massImbalanceBound} we get:
\begin{align}\label{globalMassImbalance}
\| N_T - N_0 \|_{L^1} 
= 
\left\| \sum_{j=0}^{K-1} \bigl(N_{(j+1)\Delta t} - N_{j\Delta t}\bigr) \right\|_{L^2}
\leq 
\sum_{j=0}^{K-1} \left\|\bigl(N_{(j+1)\Delta t} - N_{j\Delta t}\bigr)\right\|_{L^2}
\leq 
K\,C\,\Delta t^2 
= 
C\,T\,\Delta t.
\end{align}
Hence, the semi-implicit discretization \eqref{timeDisc} results in a global mass error that depends linearly on the time-step size $\Delta t$ and on the length of the simulation time interval. Consequently, \eqref{timeDisc} may be inappropriate when large time-steps or long horizons are needed.

\par Alternatively, discretizing the $s$ equation in \eqref{timeDiscSemiImp} (while leaving the $i$ and $r$ equations in \eqref{timeDiscSemiImp} unchanged) as:
\begin{align}\label{sTermExpl}
\left( \frac{ s_{t+1}-s_t }{ \Delta t },v\right) 
= 
-\beta \left(\frac{ s_t\, i_{t+1} }{N_t},v\right),
\end{align}
or the $i$ equation (while leaving the $s$ and $r$ equations in \eqref{timeDiscSemiImp} unchanged) as:
\begin{align}\label{iTermExpl}
\left( \frac{ i_{t+1}-i_t }{ \Delta t },w\right) 
= 
\beta \left(\frac{ i_t\, s_{t+1} }{N_t},w\right) 
- \phi_i (i_{t+1},w) 
- \nu_i (\nabla i_{t+1},\nabla w) 
+ \mu_i (i_{t+1} \,\nabla s_t ,\nabla w),
\end{align}
restores mass conservation. However, these choices introduce explicitness into the $s$ or $i$ equation, thus requiring sufficiently small $\Delta t$ to maintain stability.

\par In order to ensure both stability and mass conservation, while avoiding limitations on $\Delta t$, a \textit{fully implicit} method that solves a nonlinear system at each time-step is necessary. One example is:

\noindent \textbf{Problem 4, fully implicit time discretization:} 
\textit{Find $(s_{t+1}^{k+1}, i_{t+1}^{k+1}, r_{t+1}^{k+1})$ in $\overline{V}\times\overline{V}\times\overline{V}$ such that for all $(v,w,l)$ in $\overline{V}\times\overline{V}\times\overline{V}$:} 
\begin{align}
\begin{split}\label{implTimeDisc}
\left( \frac{ s_{t+1}^{k+1}-s_t }{ \Delta t },v\right) 
&= 
-\beta \left(\frac{ s_{t+1}^{k+1}\, i_{t+1}^k }{N_{t+1}^k},v\right),\\
\left( \frac{ i_{t+1}^{k+1}-i_t }{ \Delta t },w\right) 
&= 
\beta \left(\frac{ i_{t+1}^{k+1}\, s_{t+1}^k }{N_{t+1}^k},w\right) 
- \phi_i (i_{t+1}^{k+1},w) 
- \nu_i (\nabla i_{t+1}^{k+1},\nabla w) 
+ \mu_i (\,i_{t+1}^{k+1}\,\nabla s_{t+1}^k ,\nabla w),\\
\left( \frac{ r_{t+1}^{k+1}-r_t }{ \Delta t },l\right) 
&= 
\phi_i (i_{t+1}^{k+1},l).
\end{split}
\end{align}
The iteration is terminated upon reaching a suitable convergence criterion, then advanced to the next time level.

\par \textbf{Numerical stabilization in space.}

Sharp gradients in $s$ may cause numerical instability or spurious oscillations in solutions of \eqref{varFormS}-\eqref{varFormR}. In fact, such terms arise due to convection-like dynamics. To see this, observe that (assuming $i<<C_0$ and hence $\chi(i)\approx i$):
\begin{equation}
    \nabla \cdot \left(\mu_i i \nabla s\right) = \mu_i \nabla i \cdot \nabla s + \mu_i i \Delta s = \mu_i \left( \mathbf{u} \cdot \nabla\right)i + \mu_i i \Delta s, 
\end{equation}
where $\mathbf{u}=\nabla s$. Hence, instabilities from this term may be mitigated with  a \textit{streamline-diffusion} type stabilization. Let $\Omega_h$ be a mesh of $\Omega$ with $P$ elements, and let $\overline{V}_h \subset \overline{V}$ be a corresponding discrete subspace. Define the stabilization term $\tau$ by
\begin{equation}\label{sdScheme}
    \tau(u_h ,\,v_h,\,w_h) 
    =  
    \sum_{p=1}^{P} \dfrac{h_p}{|\nabla u_h|_{p}} 
    \bigl( \nabla u_h \cdot \nabla v_h, \,\nabla u_h \cdot \nabla w_h\bigr),
\end{equation}
where $h_p$ is the size of the $p$-th element in $\Omega_h$, and $|\nabla u_h|_p$ the maximum absolute value of $\nabla u_h$ in that element.

\par Applying \eqref{sdScheme} to \eqref{implTimeDisc} yields the following stabilized problem:

\noindent \textbf{Problem 5, fully implicit in time, stabilized in space:} 
\textit{Find $(s_{h,t+1}^{k+1}, i_{h,t+1}^{k+1}, r_{h,t+1}^{k+1})$ in $\overline{V}_h \times \overline{V}_h \times \overline{V}_h$ such that for all $(v_h,w_h,l_h)$ in $\overline{V}_h \times \overline{V}_h \times \overline{V}_h$:} 
\begin{align}
\begin{split}\label{implTimeDiscStab}
\left( \frac{ s_{h,t+1}^{k+1} - s_{t,h} }{ \Delta t },v_h\right) 
&= 
-\beta \left(\frac{ s_{h,t+1}^{k+1}\, i_{h,t+1}^k }{N_{h,t+1}^k},v_h\right),\\
\left( \frac{ i_{h,t+1}^{k+1} - i_{t,h} }{ \Delta t },w_h\right) 
&= 
\beta \left(\frac{ i_{h,t+1}^{k+1}\, s_{h,t+1}^k }{N_{h,t+1}^k},w_h\right) 
- \phi_i (i_{h,t+1}^{k+1},w_h) 
- \nu_i (\nabla i_{h,t+1}^{k+1},\nabla w_h)\\
&\qquad + \mu_i \bigl(i_{h,t+1}^{k+1}\,\nabla s_{h,t+1}^k ,\nabla w_h\bigr) 
+ \tau\bigl(s_{h,t+1}^k,\, i_{h,t+1}^{k+1},\, w_h\bigr),\\
\left( \frac{ r_{h,t+1}^{k+1} - r_{t,h} }{ \Delta t },l_h\right) 
&= 
\phi_i \bigl(i_{h,t+1}^{k+1},l_h\bigr).
\end{split}
\end{align}

It is straightforward to verify that \eqref{implTimeDiscStab} is consistent with \eqref{implTimeDisc}, since $\tau(s_{h,t+1}^k,\,i_{h,t+1}^{k+1},\,w_h)\to 0$ as $h \to 0$. However, the stabilization introduces an $\mathcal{O}(h)$ error in space. For further details on streamline-diffusion and related methods, see \cite{quarteroni2008numerical, hughes1982streamline, hughes1979finite}.

\section{Numerical examples}
In this section, we provide a series of numerical examples. Specifically, we present two such examples:
\begin{enumerate}
    \item \textit{Lombardy, early stages of COVID-19:} Following \cite{viguerie2020diffusion, viguerie2021simulating, grave2022modeling}, in this example, we simulate the Italian region of Lombardy in the early stages of the COVID-19 pandemic. This example seeks to verify whether the proposed model, with time-independent parameters, can qualitatively capture the same spatiotemporal patterns observed during the early stages of the COVID-19 pandemic. 
    \item \textit{US State of Georgia, early stages of COVID-19: } In the work \cite{grave2021assessing}, the transmission of COVID-19 in the US state of Georgia was studied with mixed results. Considering the entire state, a diffusion-based model, with time-varying diffusion and contact terms, showed good agreement with the surveillance data. However, at the local level, individual counties diverged substantially from the observed dynamics. In this example, we aim to show that the chemotaxis-inspired model produces more consistent results with reality, even when considering time-static parameters. We note that Georgia is a useful example as its county delineation provides a relatively high level of spatial resolution; its 159 counties are second only to Texas' 254; however, Georgia's counties are, on average, less than half as large (362 square kilometers compared to 1029) \cite{USDOT_BTS_Counties}.  
\end{enumerate}

We note that the current model \textit{is not intended to be a model of COVID-19.} Given the complex, heterogeneous risk levels which vary according to population characteristics, as well as the multi-stage nature of the disease, a rigorous treatment of COVID-19 requires a more complex compartmental structure than the simple $SIR$ formulation used herein \cite{jordan2020covid}. Furthermore, the large uncertainty in the surveillance data makes it challenging to directly compare a mathematical model with measured data. These examples are, instead, intended to provide a qualitative proof-of-concept; in particular, we aim to assess whether introducing a chemotaxis-like process may provide a better description of airborne infectious disease transmission, particularly regarding the spatiotemporal transmission patterns. Given the recency of the COVID-19 epidemic and the unprecedented availability of data (even if these are uncertain), this provides a natural example against which we may evaluate our model. However, accurate modeling of COVID-19 requires additional considerations, including population age structure, vaccination, and masking/mobility policies, to name only a few. For this reason, we only compare our model results with the COVID-19 data in a general, qualitative manner. Adapting the proposed model to a realistic model of COVID-19 would, however, require many additional steps beyond the focus of the current work.

\par Both simulations were performed using the Finite Element Method (FEM) for the spatial discretization with piecewise-linear elements (see, e.g., \cite{hughes2012finite, quarteroni2008numerical} for more details) and employed the same adaptive meshing scheme used in  \cite{grave2021adaptive, grave2021assessing}. Following the results in \cite{viguerie2022coupled}, the BDF2 method was used for the temporal discretization (see, e.g., \cite{quarteroni2008numerical}). In both examples, numerical stabilization was employed due to the high Peclet numbers observed for certain values of $\mu_i$; we employed the streamline-diffusion method \cite{quarteroni2008numerical, hughes1982streamline}. All simulations were performed in libMesh \cite{kirk2006libmesh}. Geometric reconstruction and meshing were done following the workflow outlined in \cite{grave2021assessing}.

\subsection{Lombardy }
\begin{figure}[ht!]
    \centering
    \includegraphics[width=.5\textwidth]{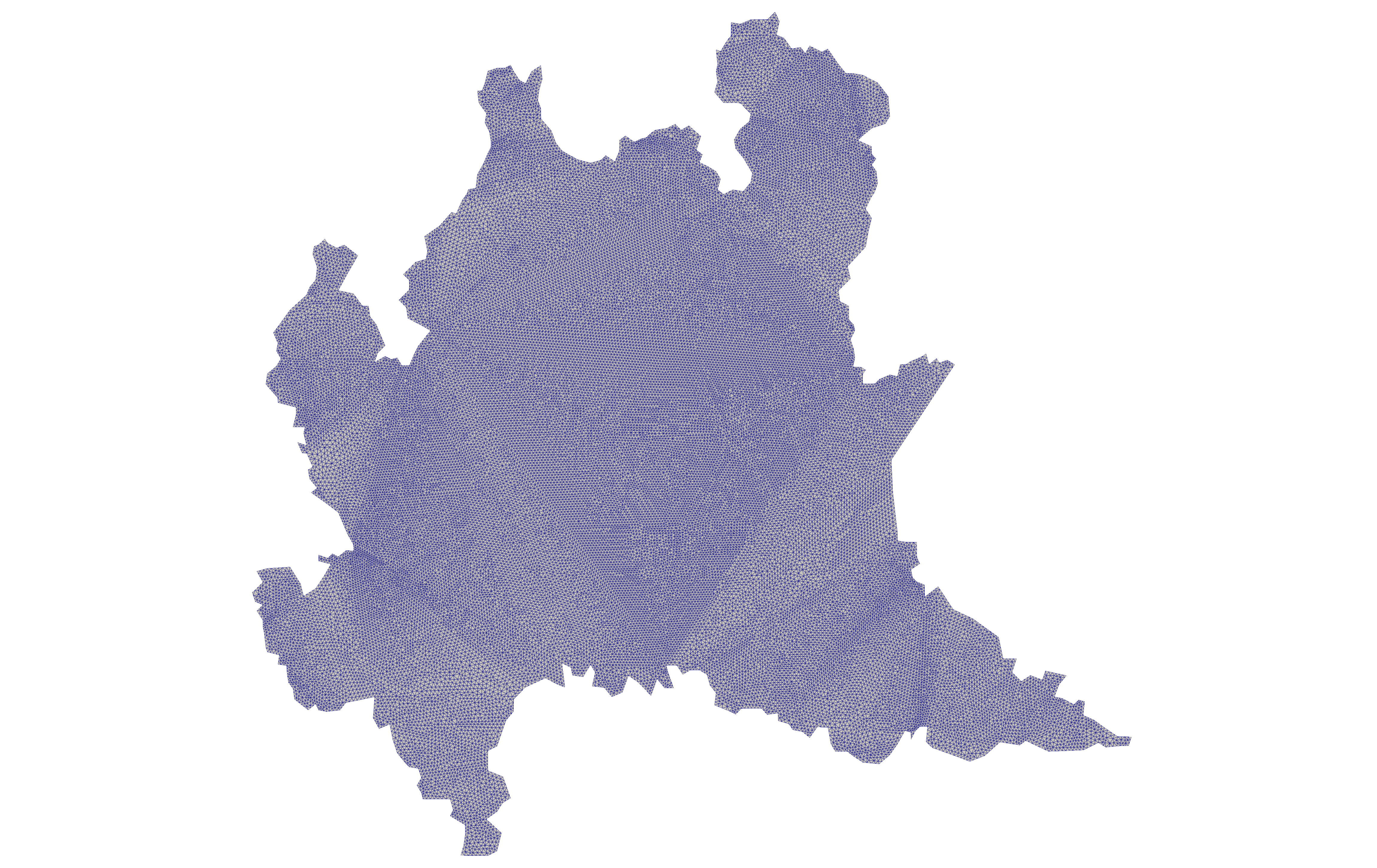}

    \caption{
    Initial mesh configuration for Lombardy simulation. 
    }
    \label{fig:lombMesh}
\end{figure}

\begin{figure}
    \centering
    \includegraphics[width=\textwidth]{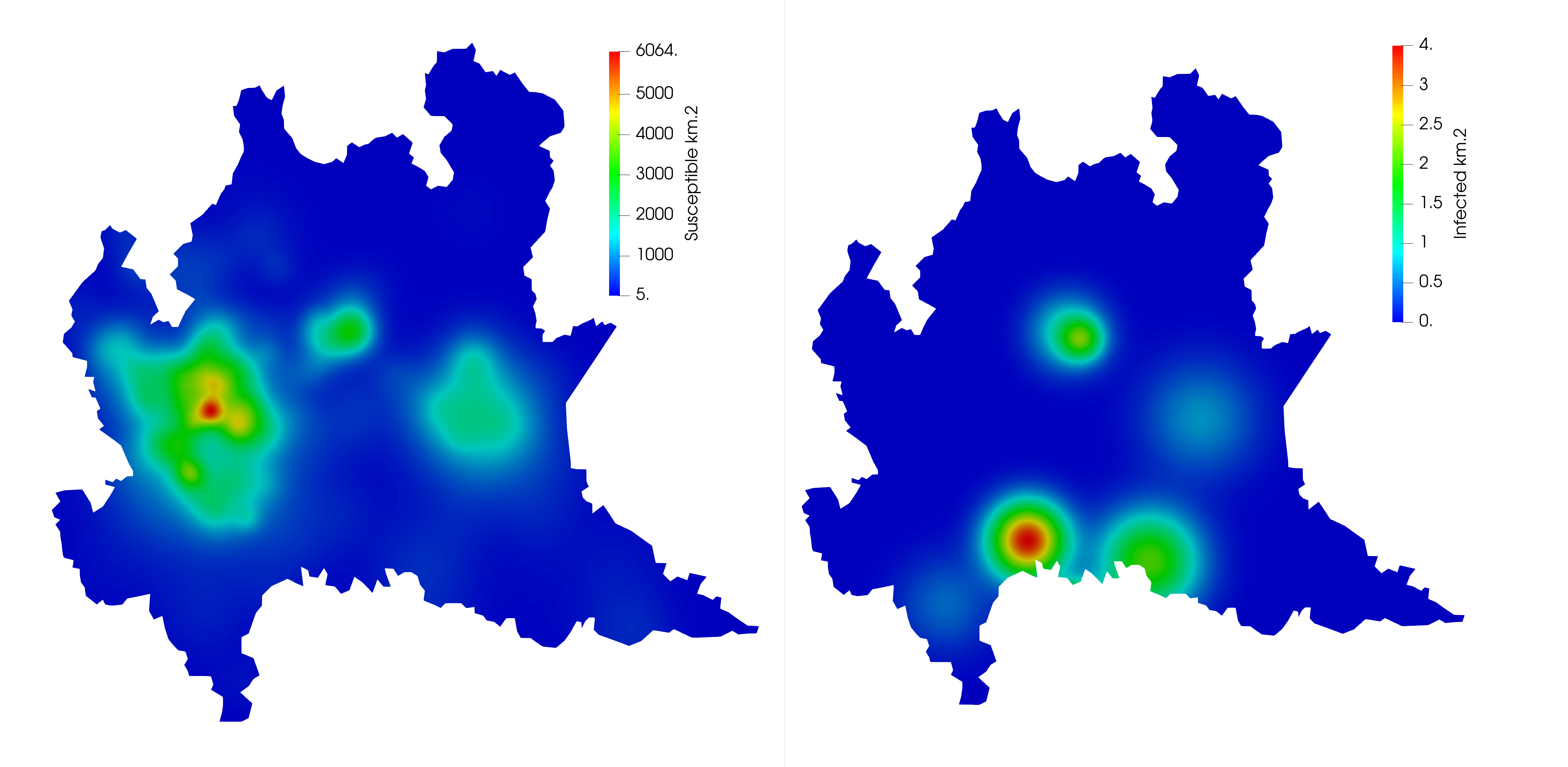}
    \caption{Initial conditions for susceptible (left) and infected (center) compartments, Lombardy simulation. Location of important provinces for the simulation study (right). }
    \label{fig:lombICs}
\end{figure}

\begin{figure}[ht!]
    \centering
    \includegraphics[width=.7\textwidth]{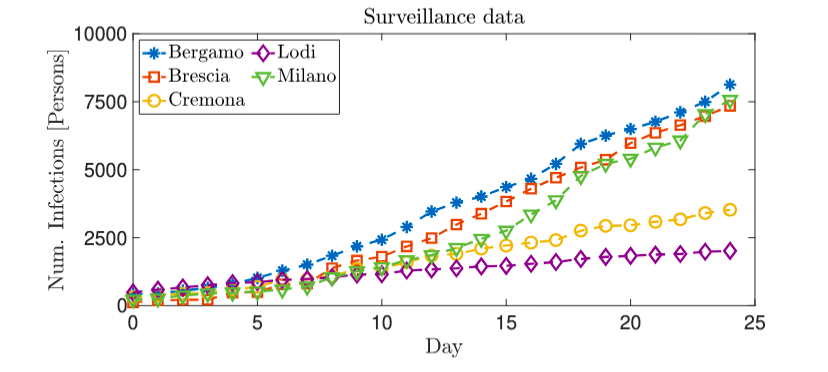}
    \includegraphics[width=.25\textwidth]{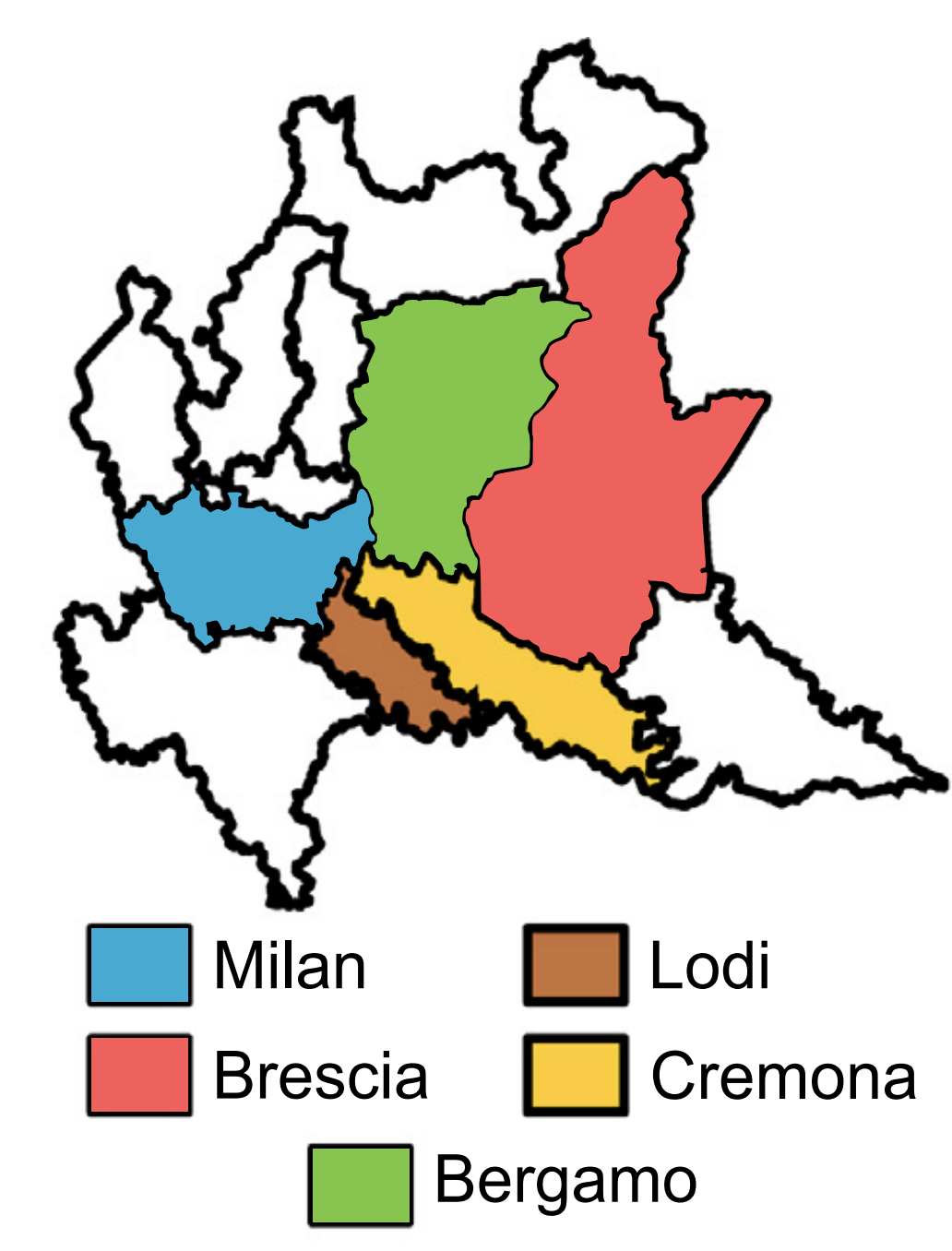}
    \caption{Left: observed cumulative infections across five relevant provinces in Lombardy (data from \cite{Lab24}). Right: geographic locations of each province.  }
    \label{fig:map_and_survLomb}
\end{figure}

\subsubsection{Simulation setup}
In Fig. \ref{fig:lombMesh}, we show the generated grid for Lombardy. The original mesh has 101,360 elements (an average resolution of 0.23 km$^2$ per element) and, after refinement, 86,038 elements. We set the time-step $\Delta t=0.25$ days and performed the adaptive mesh refinement every 16 time-steps, considering a total of 25 days. The initial susceptible and infected populations were defined as in \cite{grave2021adaptive, viguerie2021simulating, viguerie2020diffusion}, and are depicted in Figure \ref{fig:lombICs}. For all simulations, the contact rate $\beta=0.175$ / days and the recovery rate $\phi=1/18$ days.  


\subsubsection{Study design} 
We seek to assess whether the chemotaxis-inspired model produces results more consistent with observed epidemic behavior than a purely diffusive model. To this end, we performed a series of eight numerical simulations over a range of diffusion coefficients $\nu_i$ and chemotaxis parameters $\mu_i$ to examine the effect of each parameter on the subsequent spatiotemporal epidemic progression. The parameters for each simulation are given in Table \ref{tab:lombParameters}. As mentioned, we compare our simulation results qualitatively to the surveillance data collected from \cite{Lab24}, and shown in Fig. \ref{fig:map_and_survLomb} (left). In particular, we focus on the behavior of the epidemic in the provinces of Milan, Bergamo, Brescia, Cremona, and Lodi as we examine how well the simulations capture the observed spatial distribution of the epidemic over time. These data and the corresponding geographic locations are shown in Fig. \ref{fig:map_and_survLomb}.

\begin{table}
\begin{center}
\begin{tabular}{|p{.6in}|p{1.25in}|p{.7in}|p{.8in}|p{.7in}|}
\hline
Simulation & $\mu_i$ & $\nu_i$ & $\beta$ & $\phi$ \\ \hline\hline
1 &  0.0 $\dfrac{\text{km}^2 \cdot \text{Persons}}{\text{Days}}$ &  1.0 $\dfrac{\text{km}^2}{\text{Days}}$    & 0.175 Days$^{-1}$  & 1/18 Days$^{-1}$ \\ \hline
2 &  0.0 $\dfrac{\text{km}^2 \cdot \text{Persons}}{\text{Days}}$ &  2.5 $\dfrac{\text{km}^2}{\text{Days}}$    & 0.175 Days$^{-1}$  & 1/18 Days$^{-1}$ \\ \hline
3 &  0.01 $\dfrac{\text{km}^2 \cdot \text{Persons}}{\text{Days}}$ &  1.0 $\dfrac{\text{km}^2}{\text{Days}}$    & 0.175 Days$^{-1}$  & 1/18 Days$^{-1}$ \\ \hline
4 &  0.01 $\dfrac{\text{km}^2 \cdot \text{Persons}}{\text{Days}}$ &  2.5 $\dfrac{\text{km}^2}{\text{Days}}$    & 0.175 Days$^{-1}$  & 1/18 Days$^{-1}$ \\ \hline
5 &  0.015 $\dfrac{\text{km}^2 \cdot \text{Persons}}{\text{Days}}$ &  1.0 $\dfrac{\text{km}^2}{\text{Days}}$    & 0.175 Days$^{-1}$  & 1/18 Days$^{-1}$ \\ \hline
6 &  0.015 $\dfrac{\text{km}^2 \cdot \text{Persons}}{\text{Days}}$ &  2.5 $\dfrac{\text{km}^2}{\text{Days}}$    & 0.175 Days$^{-1}$  & 1/18 Days$^{-1}$ \\ \hline
7 &  0.02 $\dfrac{\text{km}^2 \cdot \text{Persons}}{\text{Days}}$ &  1.0 $\dfrac{\text{km}^2}{\text{Days}}$    & 0.175 Days$^{-1}$  & 1/18 Days$^{-1}$ \\ \hline
8 &  0.02 $\dfrac{\text{km}^2 \cdot \text{Persons}}{\text{Days}}$ &  2.5 $\dfrac{\text{km}^2}{\text{Days}}$    & 0.175 Days$^{-1}$  & 1/18 Days$^{-1}$ \\ \hline
\end{tabular}%
\end{center}
\caption{Parameter values for each of the eight numerical simulations for Lombardy.}\label{tab:lombParameters}
\end{table}

\subsubsection{Results}

\begin{figure}[ht!]
    \centering
    \includegraphics[width=\textwidth]{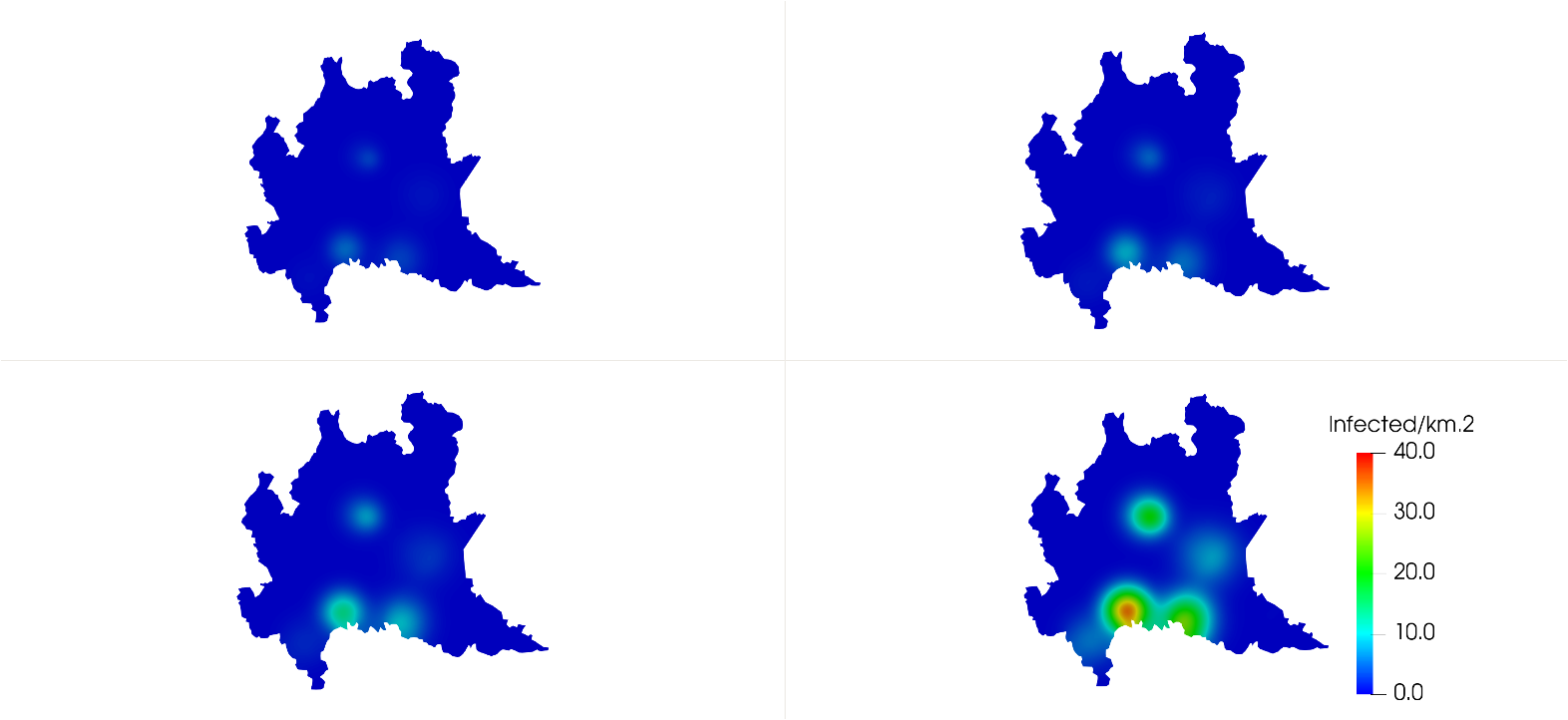}
    \caption{Days 5 (top-left), 10 (top-right), 15 (bottom-left) and 25 (bottom-right) of the Lombardy simulation for $\mu_i=0.0$ and $\nu_i=1.0$ km$^2 \cdot$ Days $^{-1}$.} 
    \label{fig:lombChem0Diff1}
\end{figure}

\begin{figure}[ht!]
    \centering
    \includegraphics[width=\textwidth]{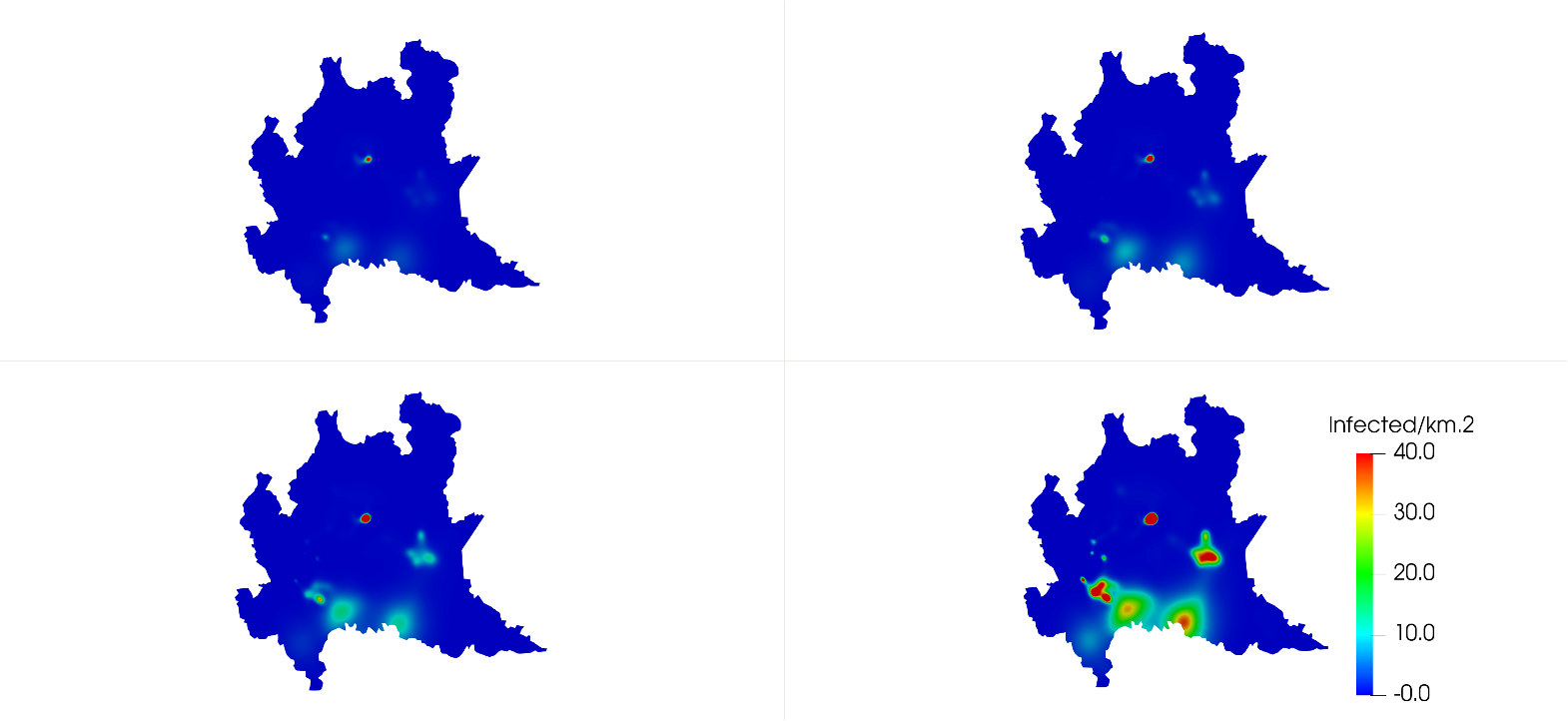}
    \caption{Days 5 (top-left), 10 (top-right), 15 (bottom-left) and 25 (bottom-right) of the Lombardy simulation for $\mu_i=0.01$ and $\nu_i=1.0$ km$^2 \cdot$ Days $^{-1}$.}
    \label{fig:lombChem1Diff1}
\end{figure}

\begin{figure}[ht!]
    \centering
    \includegraphics[width=\textwidth]{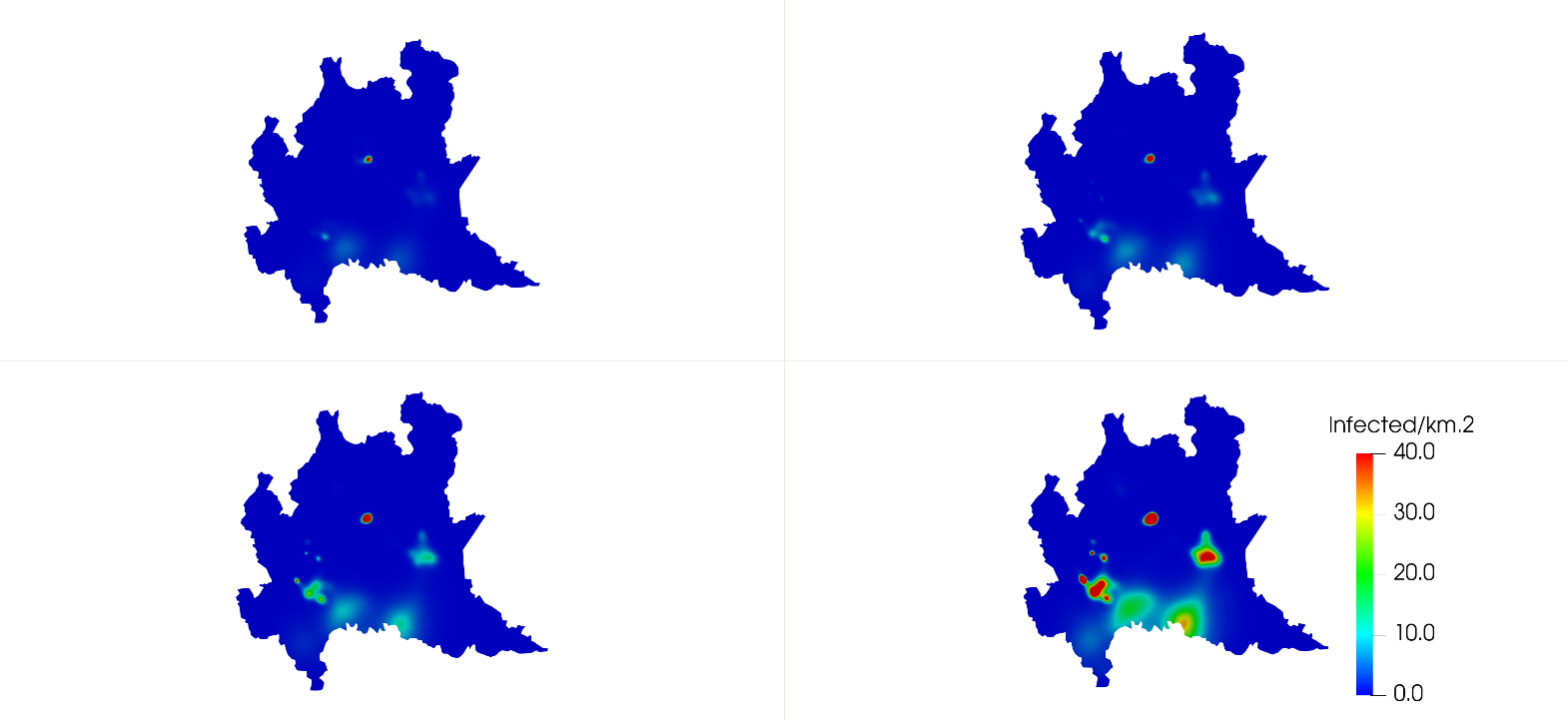}
    \caption{Days 5 (top-left), 10 (top-right), 15 (bottom-left) and 25 (bottom-right) of the Lombardy simulation for $\mu_i=0.02$ and $\nu_i=2.5$ km$^2 \cdot$ Days $^{-1}$.}
    \label{fig:lombChem2Diff25}
\end{figure}

In Fig. \ref{fig:lombChem0Diff1}-\ref{fig:lombChem2Diff25}, we show the infected compartment $i$ at days 5, 10, 15, and 25 for the simulation studies with $\mu_i=0.00$, $\nu_i=1.0$, $\mu_i=0.01$, $\nu_i=1.0$, and $\mu_i=0.02$, $\nu_i=2.5$, respectively. When comparing with the initial conditions depicted in Fig. \ref{fig:lombICs}, we see that the case with no chemotaxis (Fig. \ref{fig:lombChem0Diff1}) shows little qualitative difference. The epidemic remains primarily concentrated in the southern provinces of Lodi and Cremona and the surrounding areas, with lesser contagion observed in the regions of Brescia and Bergamo. Very little contagion is observed in the Milan province and the surrounding areas. 

\par However, when the chemotaxis parameter is introduced, we immediately observe significant qualitative differences in the spatiotemporal dynamics. Observing Fig. \ref{fig:lombChem1Diff1}, while the epidemic does continue in the Lodi and Cremona regions, contagion quickly becomes more pronounced in the  Bergamo and Brescia regions. Additionally, the epidemic front moves northwest from Lodi into Milan and the surrounding areas. By day 25, the contagion progresses into several densely populated regions north of Milan. In Fig. \ref{fig:lombChem2Diff25}, we see that further increases in the chemotaxis parameter result in greater epidemic intensity in Milan and its surrounding areas. In particular, intense outbreaks occur in several northern suburbs around the city center. 

\par
\begin{figure}[ht!]
    \centering
    \includegraphics[width=\textwidth]{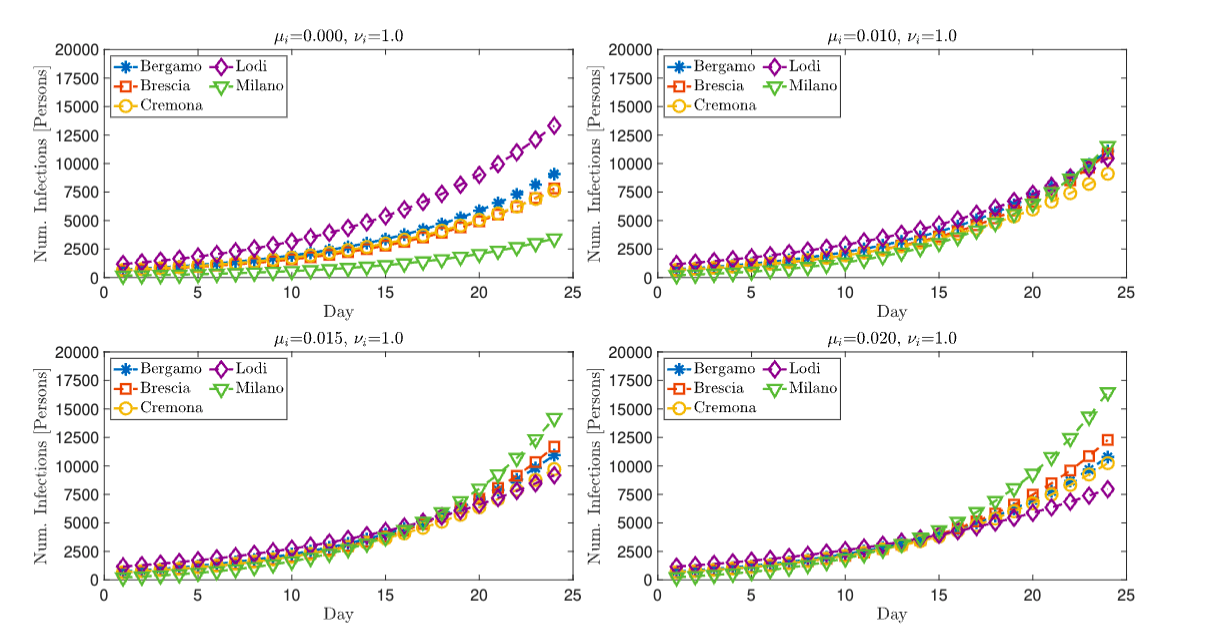}
    \caption{Lombardy study: cumulative incidence across five different provinces for $\nu_i=1.0$ km$^{2} \cdot$ Days $^{-1}$ and varying levels of the chemotaxis parameter.  }
    \label{fig:lombDiff1pt0}
\end{figure}

 In Fig. \ref{fig:lombDiff1pt0}, we plot the cumulative incidence in each of the five major provinces over the simulation period for varying values of the chemotaxis parameter $\mu_i$ for all simulations with $\nu_i=1.0$ km$^{2}\cdot$ Days $^{-1}$. This plot confirms the qualitative patterns shown in Fig. \ref{fig:lombChem0Diff1}-\ref{fig:lombChem1Diff1}. When no chemotaxis is considered, the epidemic remains most concentrated in the Lodi area, with Cremona, Bergamo, and Brescia showing similar overall intensity levels. Furthermore, very little contagion is observed in the Milan area. 

\par As $\mu_i$ is increased, the Bergamo, Brescia, and Milan areas become the most heavily affected regions. For $\mu_i=0.02$, this effect is quite pronounced in Milan, and nearly twice as many infections occur in Milan compared to the next most affected region. We also note that chemotaxis increases overall contagion in Cremona. However, these increases are much smaller than those in Bergamo, Brescia, and Milan. Interestingly, however, increased chemotaxis reduces epidemic intensity in Lodi, with fewer infections for higher $\mu_i$.

\par
\begin{figure}[ht!]
    \centering
    \includegraphics[width=\textwidth]{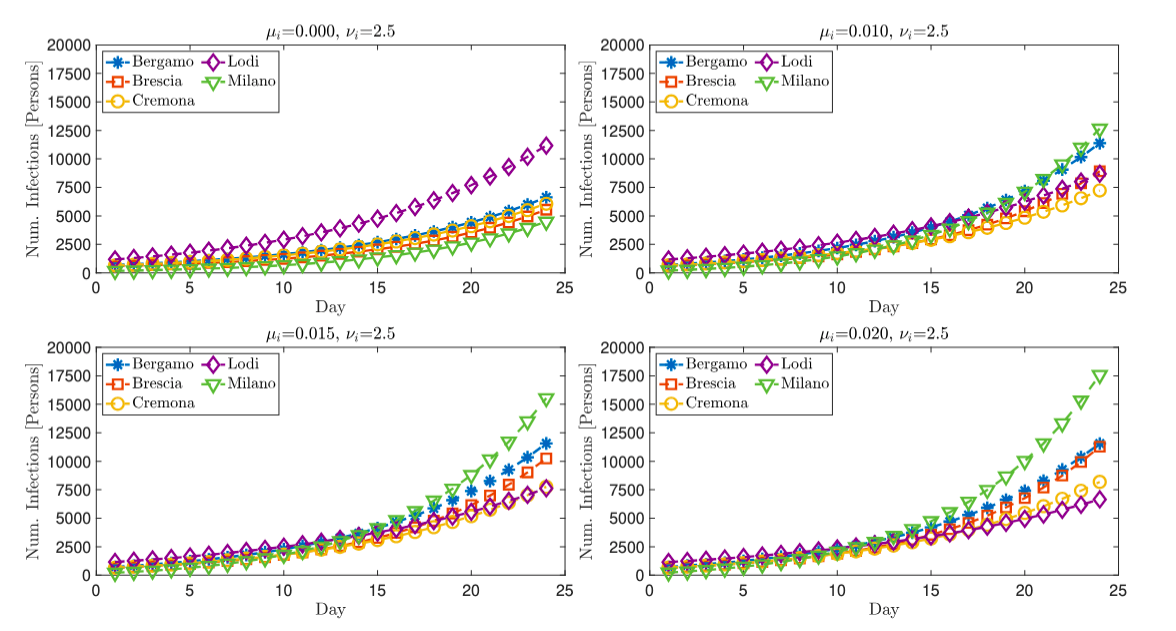}
    \caption{Lombardy study: cumulative incidence across five different provinces for $\nu_i=2.5$ km$^{2} \cdot$ Days $^{-1}$ and varying levels of the chemotaxis parameter.}
    \label{fig:lombDiff2pt5}
\end{figure}

 In Fig. \ref{fig:lombDiff2pt5}, we again plot the cumulative incidence in each of the five major provinces over the simulation period for varying chemotaxis while increasing the diffusion parameter $\nu_i=2.5$ km$^{2}\cdot$ Days $^{-1}$. Comparing Figs. \ref{fig:lombDiff1pt0} and \ref{fig:lombDiff2pt5}, we can further observe the effect of the diffusion parameter $\nu_i$ on the spatiotemporal epidemic progression. There is little qualitative difference in the pure diffusion (no chemotaxis) case compared to the corresponding simulation shown in Fig. \ref{fig:lombDiff1pt0}. When chemotaxis is increased, the effect on epidemic progression is similar for the differing levels of $\nu_i$; however, we note that increased diffusion results in somewhat higher transmission in Bergamo and somewhat lower transmission in Brescia.

\par Comparing the results in \ref{fig:lombDiff1pt0}-\ref{fig:lombDiff2pt5} to the surveillance data shown in \ref{fig:map_and_survLomb} suggests that the spatiotemporal epidemic progression produced by the chemotaxis model is more consistent with the observed data. The surveillance data shows that the epidemic was initially most concentrated in the Lodi province, with similar levels of contagion observed in Cremona, Bergamo, and Brescia and nearly zero cases in Milan. However, over the course of several weeks, the epidemic grew rapidly in the major urban centers of Bergamo, Brescia, and Milan, while growth in Lodi and Cremona was comparatively slower. 

\par
\begin{figure}[ht!]
    \centering
    \includegraphics[width=\textwidth]{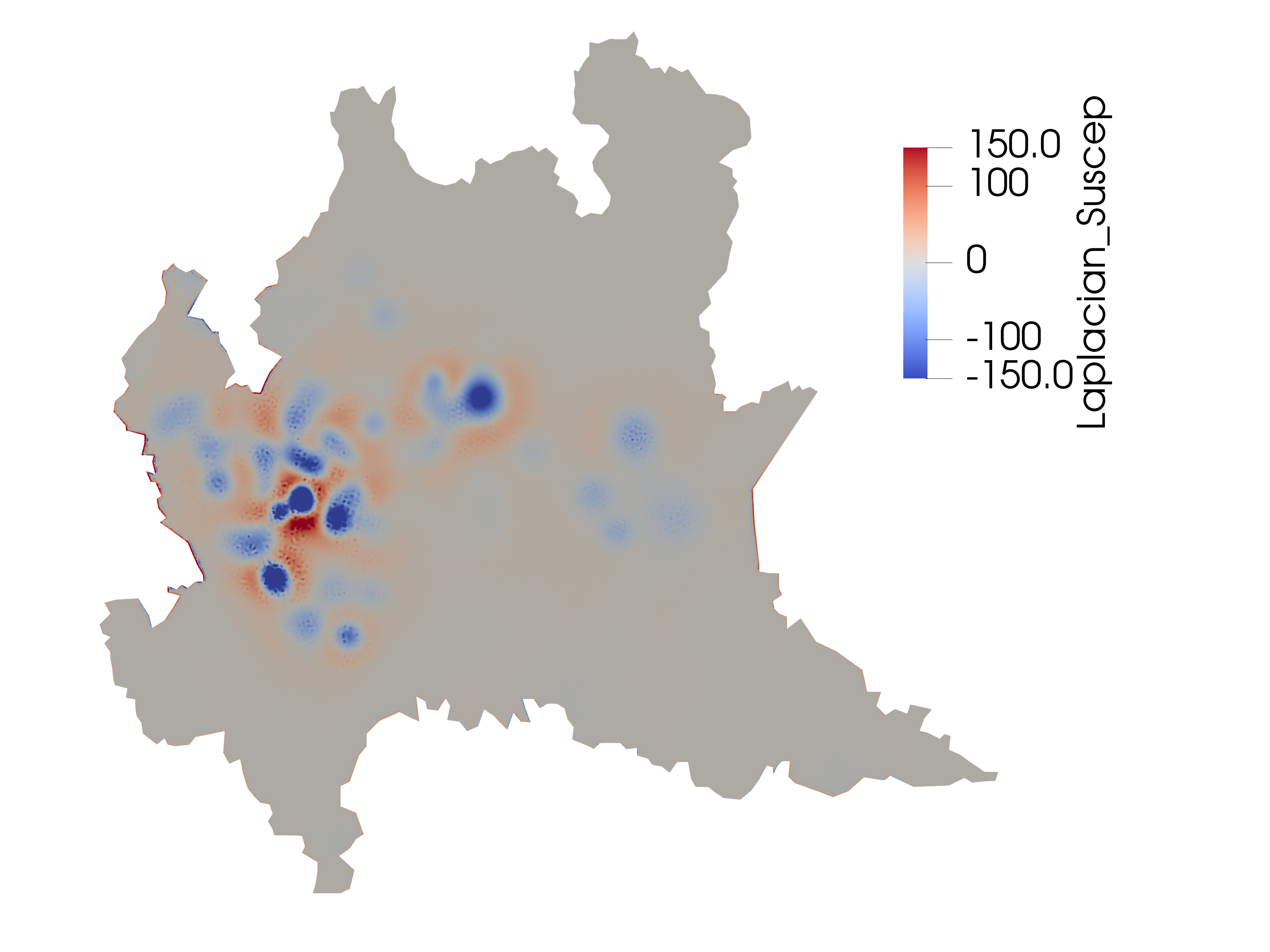}
    \caption{The Laplacian $\Delta s$ of the initial susceptible population $s$ in Lombardy. Large negative values indicate areas particularly vulnerable to the epidemic. }
    \label{fig:lombLaplace}
\end{figure}

 A purely diffusive model cannot capture these important dynamics; the Lodi area remains the most heavily affected region after 25 days, and the Milan area shows essentially no epidemic activity. However, when chemotaxis is introduced, the model behavior is more consistent with the surveillance data. In particular, the epidemic grows faster in Bergamo, Brescia, and Milan than in the southern regions. Furthermore, the epidemic model can also replicate infection hotspots or areas of particularly high transmission. A notable example of this phenomenon is the city of Bergamo, as can be observed in Figs. \ref{fig:lombChem1Diff1} and \ref{fig:lombChem2Diff25}, a region of sustained, high transmission develops around this area. This hotspot replication is consistent with reality, and this area was indeed particularly hard-hit in the early stages of the COVID-19 epidemic \cite{Lab24}. The development of a hotspot in Bergamo is well-supported by the theory of the chemotaxis model. In Fig. \ref{fig:lombLaplace}, we plot $\Delta s$ for the initial susceptible population. As observed, $\Delta s$ has a particularly large negative value around Bergamo. Indeed, the derived reproduction number \ref{spatialR0Defn} suggests that such an area may be particularly vulnerable to an outbreak. 
\par Additionally, while the chemotaxis model does not, strictly speaking, incorporate nonlocal behavior, some of the simulation results nonetheless show interesting dynamics, which resemble nonlocal behavior. In particular, when comparing figures \eqref{fig:lombChem0Diff1} and \eqref{fig:lombChem1Diff1}-\eqref{fig:lombChem2Diff25}, we observe that the purely diffusive model results in large, smoothly-varying areas of infection density. In contrast, high-transmission areas far from the primary transmission centers appear in the chemotaxis simulations. For example, three such regions can be observed in the bottom-right panel of Figure \ref{fig:lombChem1Diff1}, where three distinct, small infection centers appear in the region's western part, disconnected from the other, larger transmission areas. 

\subsection{Georgia}

\begin{figure}[ht!]
    \centering
    \includegraphics[width=.9\textwidth]{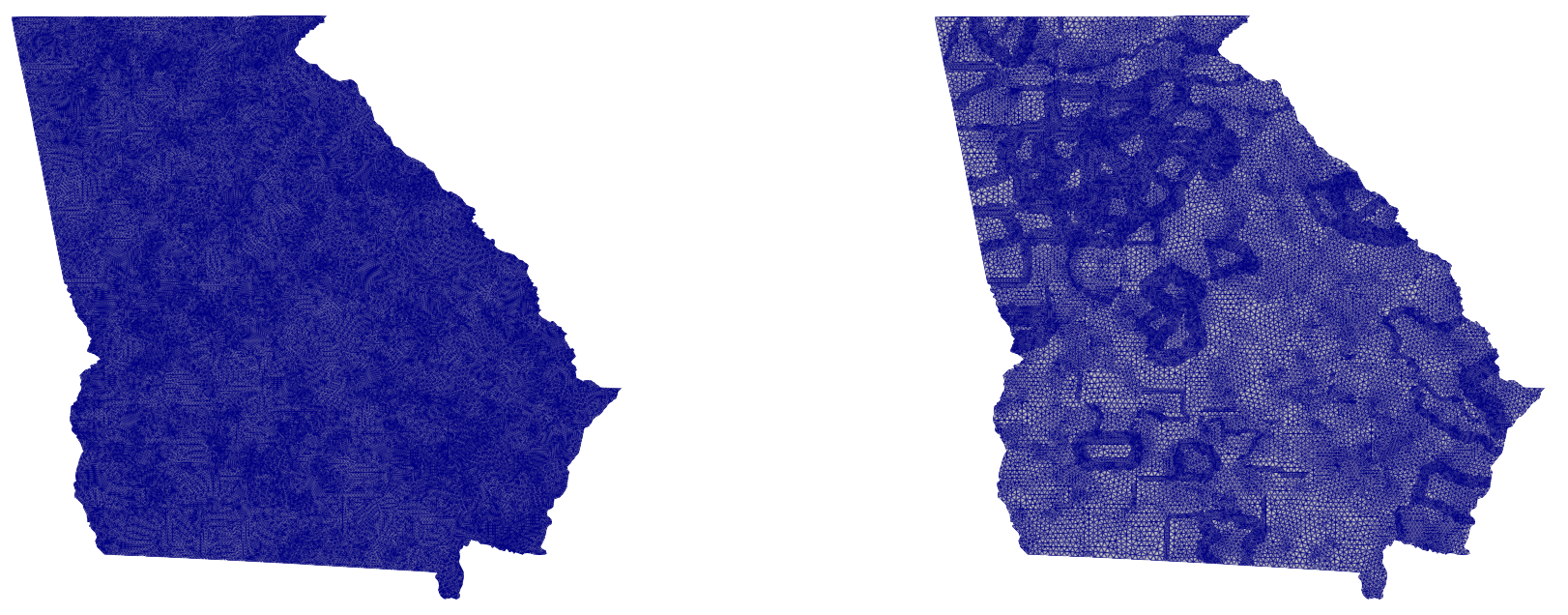}

    \caption{Left: Initial mesh configuration for Georgia simulation. Right: Final mesh configuration.}
    \label{fig:georgMesh}
\end{figure}

\begin{figure}[ht!]
    \centering
    \includegraphics[width=\textwidth]{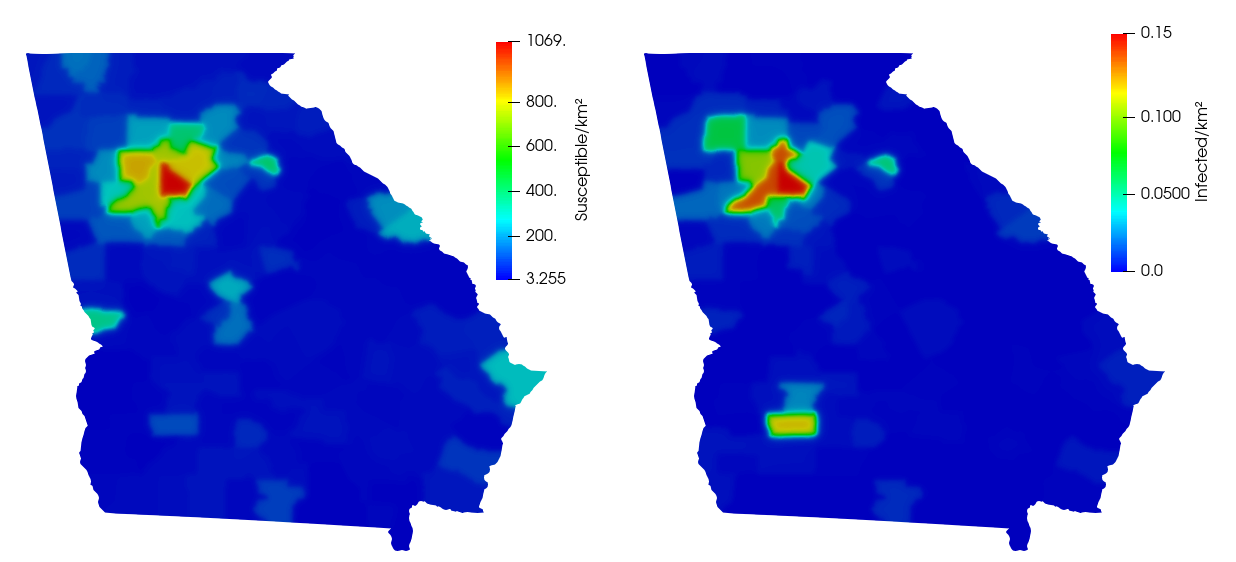}
    \caption{Initial conditions for susceptible (left) and infected (right) compartments, Georgia simulation.}
    \label{fig:georICs}
\end{figure}

\begin{figure}[ht!]
    \centering
    \includegraphics[width=.525\textwidth]{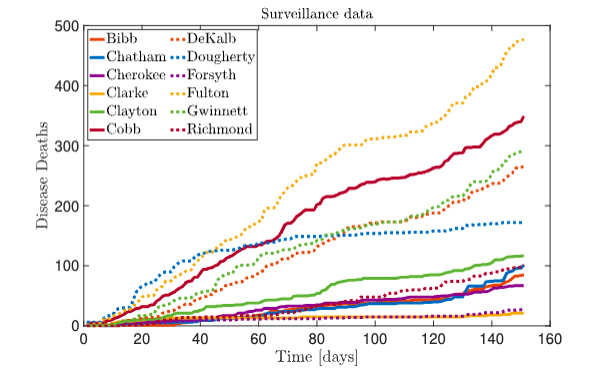}
    \includegraphics[width=.45\textwidth]{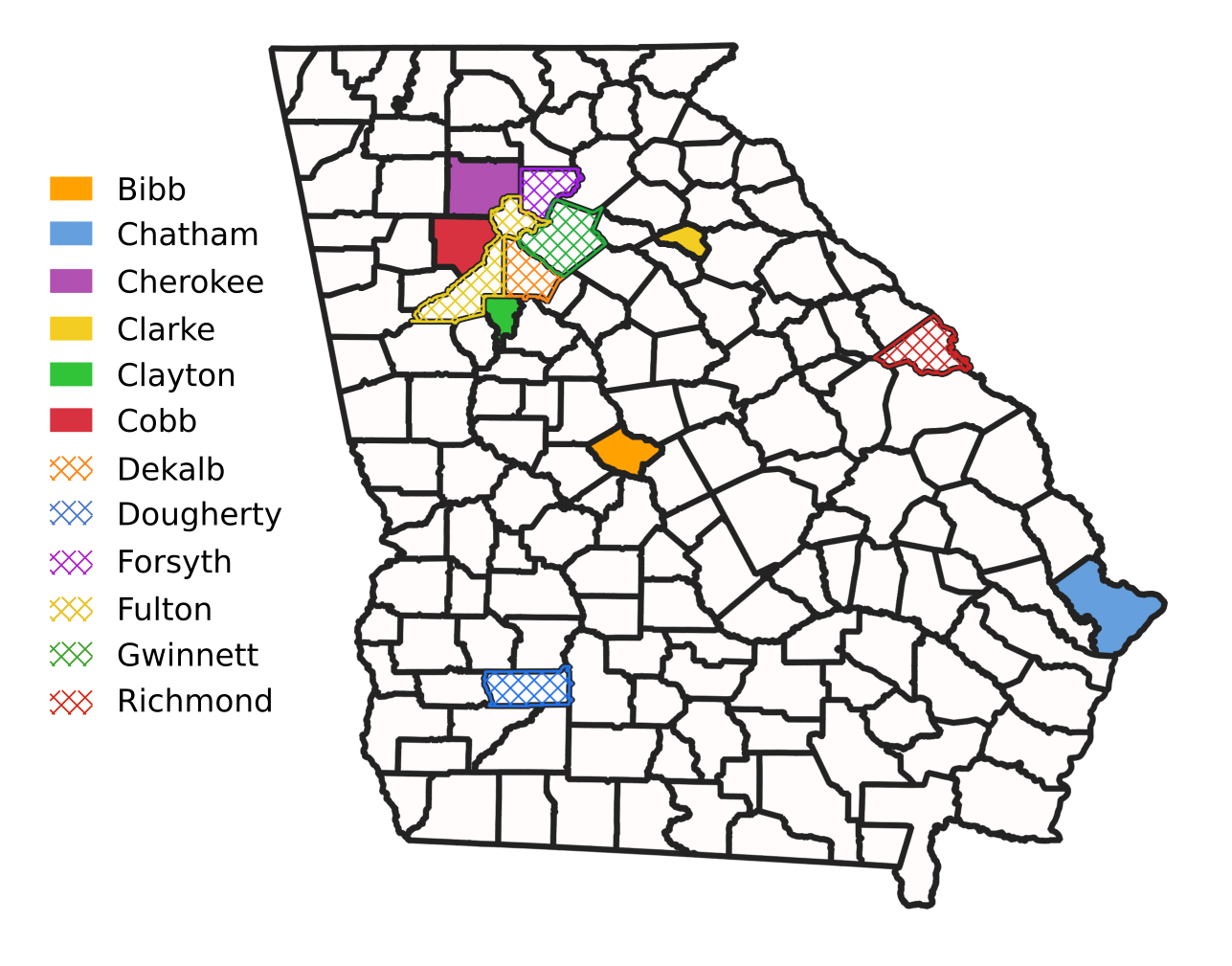}
    \caption{Left: observed COVID-19 mortality across twelve relevant counties in Georgia (data from \cite{GAUSAFacts}). Right: geographic locations of each county.}
    \label{fig:map_and_survGeorg}
\end{figure}

\subsubsection{Simulation setup}

In Fig. \ref{fig:georgMesh}, we plot the mesh used for the Georgia simulation. We used an initial triangular mesh with 128,224 elements (an average spatial resolution of 1.83 km$^2$) before refinement. The mesh was refined every 16 time-steps following the adaptive meshing procedure detailed in \cite{grave2021adaptive}; the final mesh consisted of 115,592 triangular elements. The initial conditions for the susceptible and infected are plotted in Fig. \ref{fig:georICs}.

\subsubsection{Study design}
These simulations aim to assess the ability of the purely diffusive and chemotaxis models to capture important qualitative behaviors over longer periods. As with the Lombardy simulations, we are particularly interested in the model behavior with a time-constant parameterization. In \cite{grave2021assessing}, a purely diffusive model provided reasonably accurate results for a COVID-19 simulation of the entire state of Georgia; however, this required a time-dependent parameterization. Furthermore, even with time-dependent parameters, the purely diffusive model struggled to capture certain dynamics successfully. 
\par Most importantly, while several counties in the Atlanta area (Fulton, DeKalb, Gwinnett, and Cobb) were well-captured, virtually no transmission occurred in other major population centers outside of this area (Chatham, Richmond, and Bibb counties). This problem is likely due to a combination of low levels of infection in the initial conditions, combined with the diffusive model being unable to account for the tendency of epidemics to spread preferentially to population centers \cite{sy2021population}. Ideally, the chemotaxis model will provide superior results and should increase transmission in these areas. Furthermore, other areas (particularly Forsyth and Cherokee counties) showed higher transmission levels than observed in reality. While these areas are quite close to Atlanta, they have a much lower overall population density. Therefore, we expect the chemotaxis model to reduce transmission in these areas further. 
\par   To this end, we performed three simulations: one with no chemotaxis ($\mu_i=0.0$) and two with chemotaxis ($\mu_i=0.01,\,0.02$). The remaining parameters were identical in each simulation. The complete list of parameter values for the Georgia simulations is provided in Table \ref{tab:georgiaParameters}; the differences between these and the values used in Table \ref{tab:lombParameters} may potentially be explained by differences in population mobility and/or density patterns \cite{grave2021assessing}. We compare the removed population in each simulation to the mortality data obtained from surveillance \cite{GAUSAFacts}. While we compare our simulations with COVID-19 data,  as with the Lombardy simulation, we emphasize that this comparison is purely qualitative. As dynamics important for the simulation of COVID-19, including incubation periods and population demographic considerations, were not considered, we do not believe that a quantitative comparison is appropriate in the context of the current study. 

\begin{table}
\begin{center}
\begin{tabular}{|p{.6in}|p{1.25in}|p{.7in}|p{.8in}|p{.7in}|}
\hline
Simulation & $\mu_i$ & $\nu_i$ & $\beta$ & $\phi$ \\ \hline\hline
1 &  0.0 $\dfrac{\text{km}^2 \cdot \text{Persons}}{\text{Days}}$ &  0.25 $\dfrac{\text{km}^2}{\text{Days}}$    & 0.06 Days$^{-1}$  & 1/24 Days$^{-1}$ \\ \hline
2 &  0.01 $\dfrac{\text{km}^2 \cdot \text{Persons}}{\text{Days}}$ &  0.25 $\dfrac{\text{km}^2}{\text{Days}}$    & 0.06 Days$^{-1}$  & 1/24 Days$^{-1}$ \\ \hline
3 &  0.02 $\dfrac{\text{km}^2 \cdot \text{Persons}}{\text{Days}}$ &  0.25 $\dfrac{\text{km}^2}{\text{Days}}$    & 0.06 Days$^{-1}$  & 1/24 Days$^{-1}$ \\ \hline
\end{tabular}%
\end{center}
\caption{Parameter values for the Georgia simulations.}\label{tab:georgiaParameters}
\end{table}

\subsubsection{Results}
\begin{figure}[ht!]
\centering
\begin{minipage}{.33\textwidth}
  \centering
  \includegraphics[width=.8\linewidth]{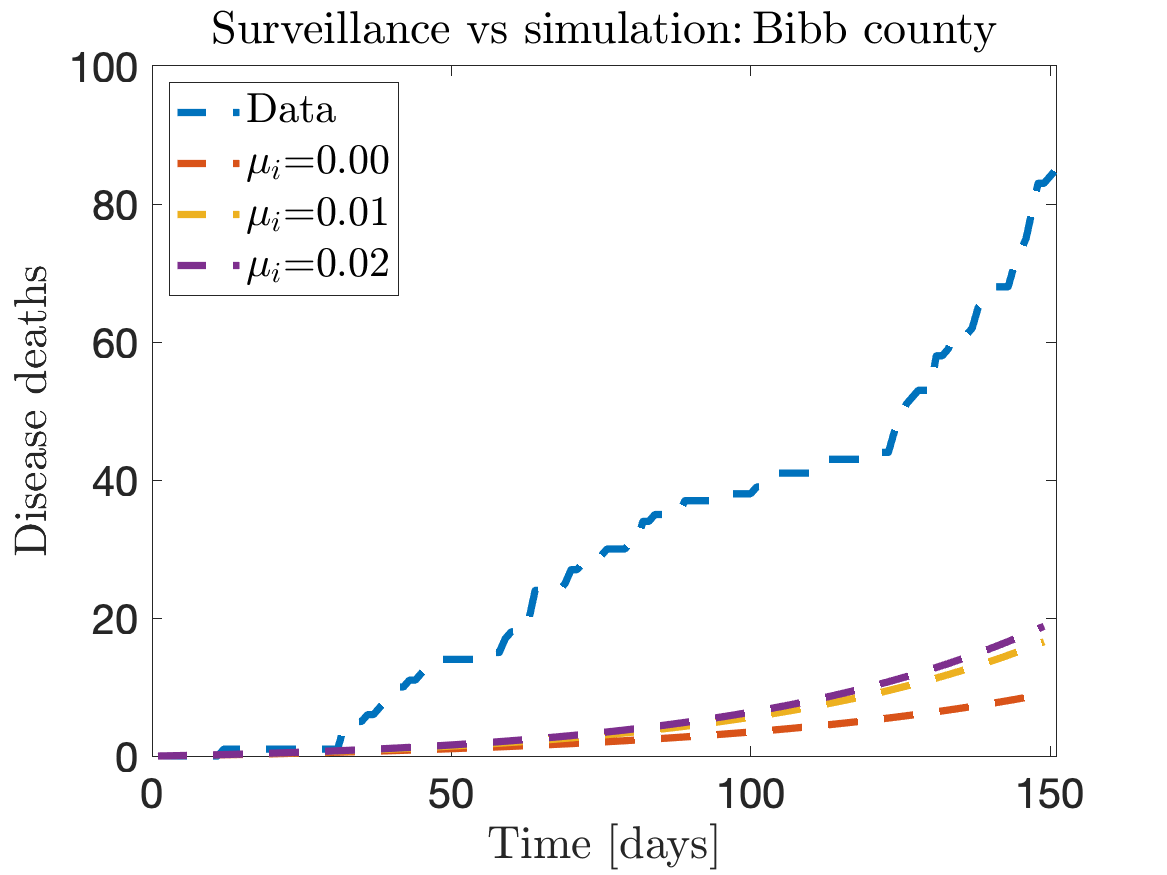}
  \caption{Georgia, Bibb County}
  \label{fig:Bibb}
\end{minipage}%
\begin{minipage}{.33\textwidth}
  \centering
  \includegraphics[width=.8\linewidth]{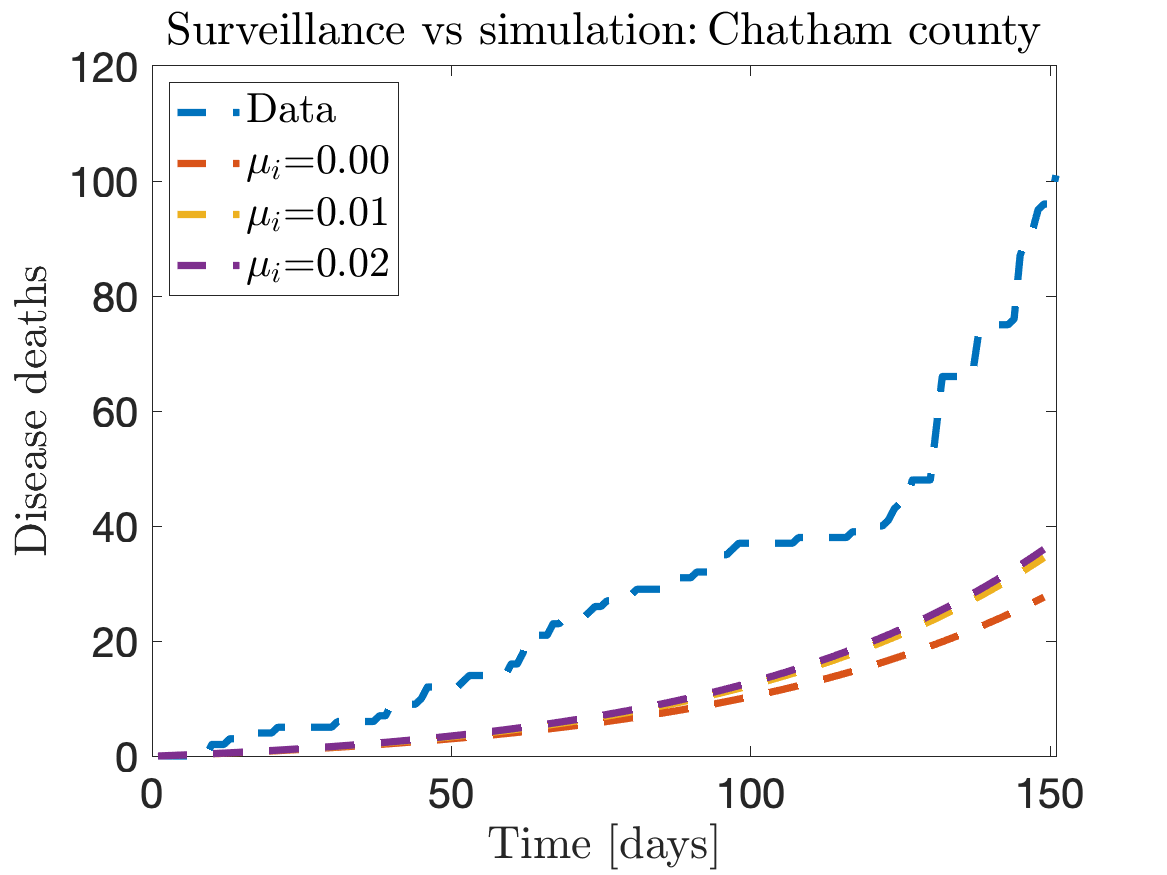}
\caption{Georgia, Chatham County}
  \label{fig:Chatham}
\end{minipage}
\begin{minipage}{.33\textwidth}
  \centering
  \includegraphics[width=.8\linewidth]{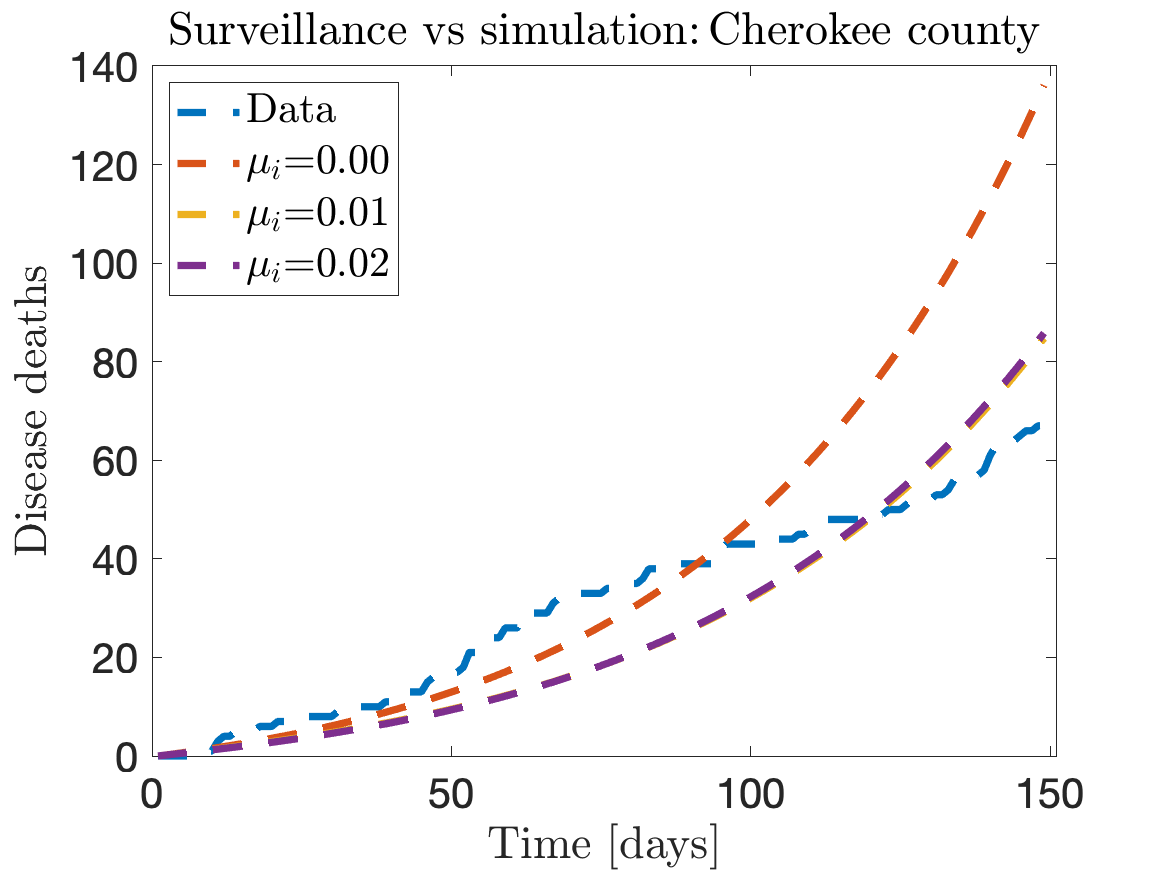}
\caption{Georgia, Cherokee County}
  \label{fig:Cherokee}
\end{minipage}
\end{figure}

\begin{figure}[ht!]
\centering
\begin{minipage}{.33\textwidth}
  \centering
  \includegraphics[width=.8\linewidth]{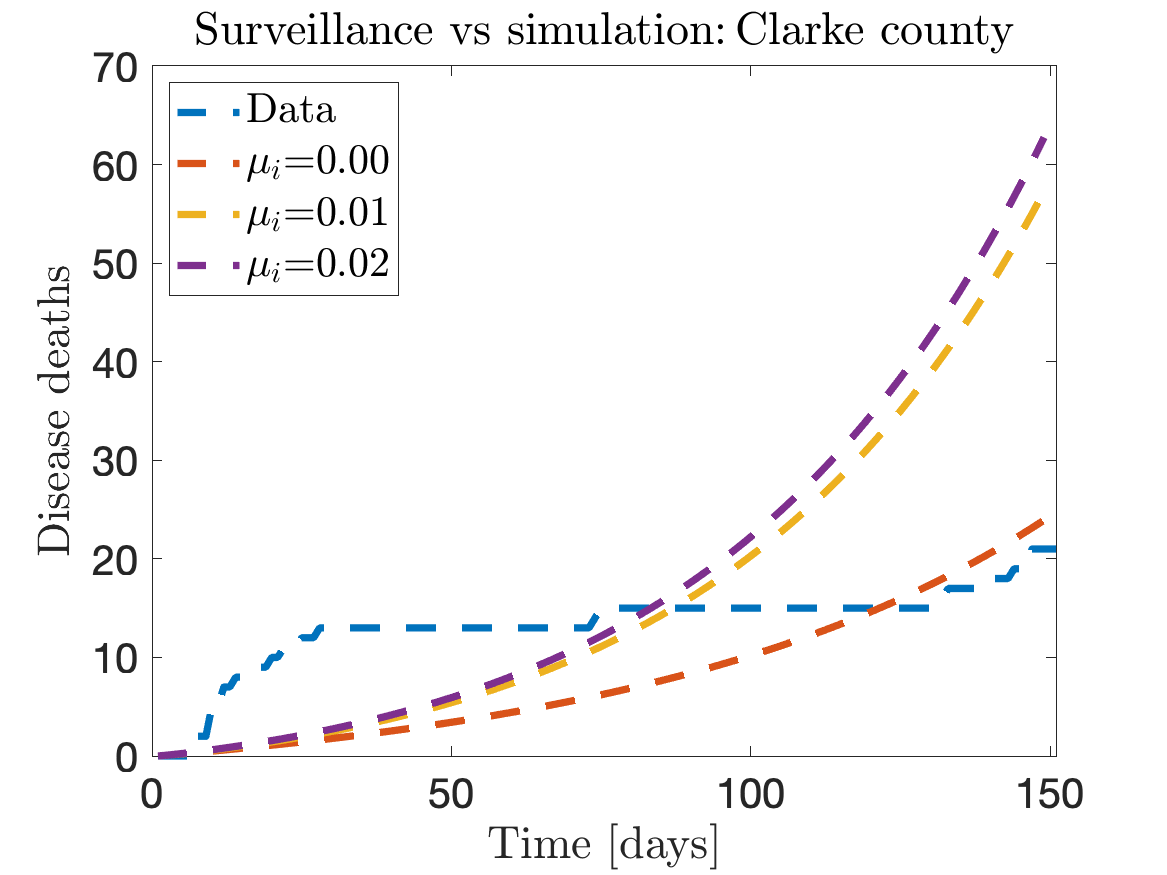}
  \caption{Georgia, Clarke County}
  \label{fig:Clarke}
\end{minipage}%
\begin{minipage}{.33\textwidth}
  \centering
  \includegraphics[width=.8\linewidth]{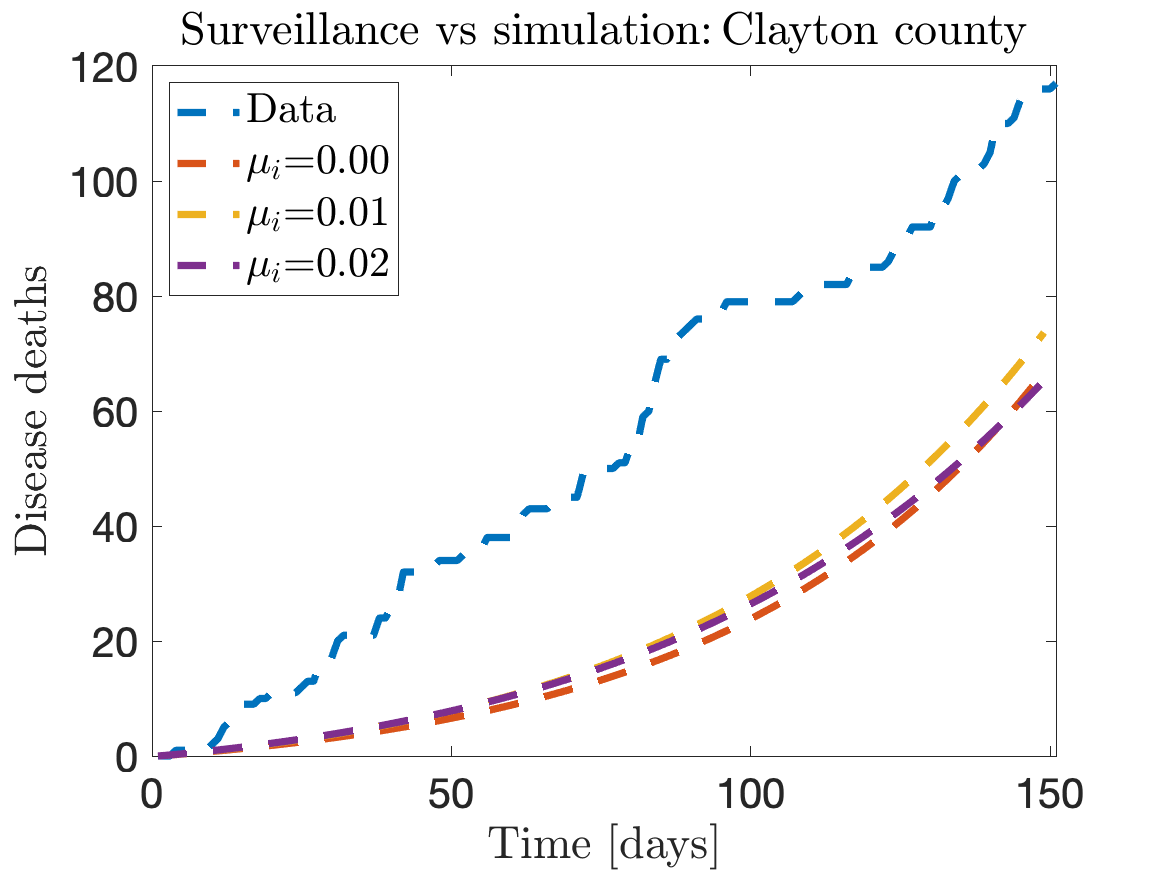}
\caption{Georgia, Clayton County}
  \label{fig:Clayton}
\end{minipage}
\begin{minipage}{.33\textwidth}
  \centering
  \includegraphics[width=.8\linewidth]{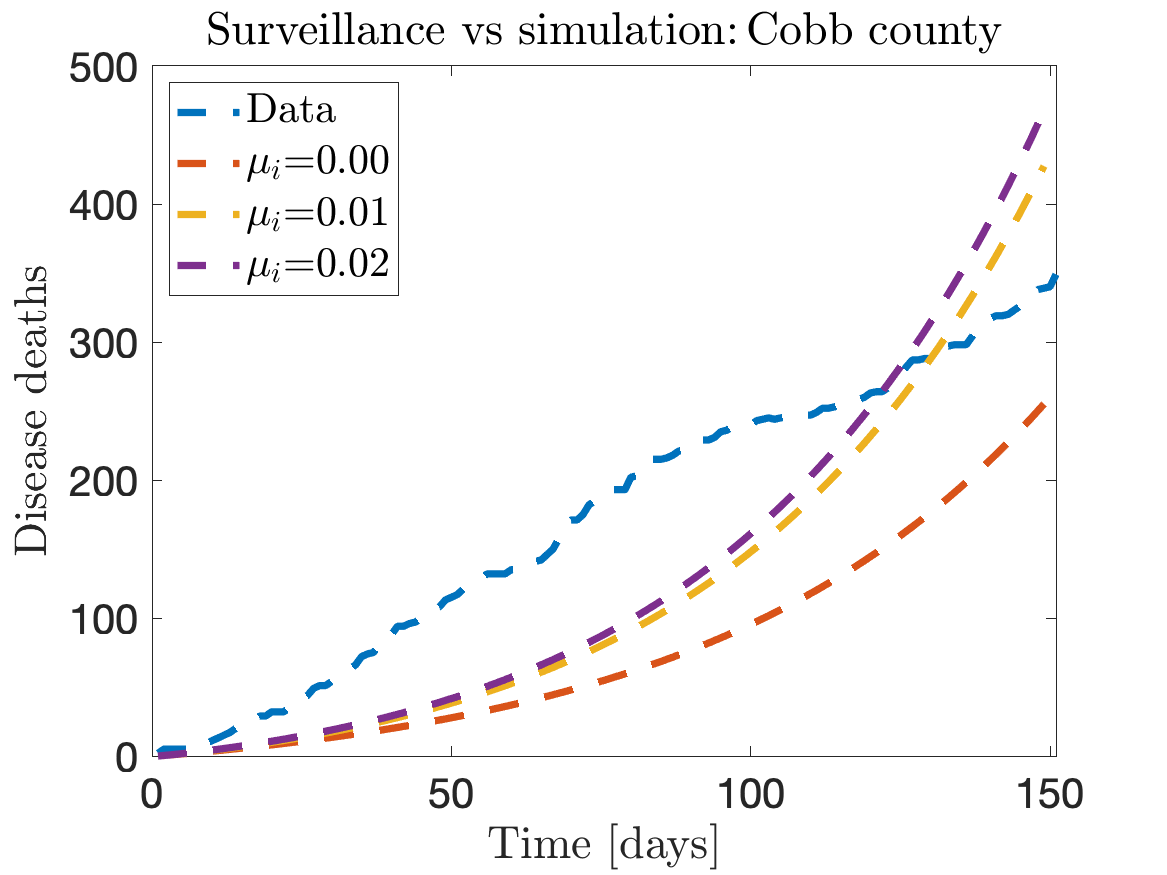}
\caption{Georgia, Cobb County}
  \label{fig:Cobb}
\end{minipage}
\end{figure}

\begin{figure}[ht!]
\centering
\begin{minipage}{.33\textwidth}
  \centering
  \includegraphics[width=.8\linewidth]{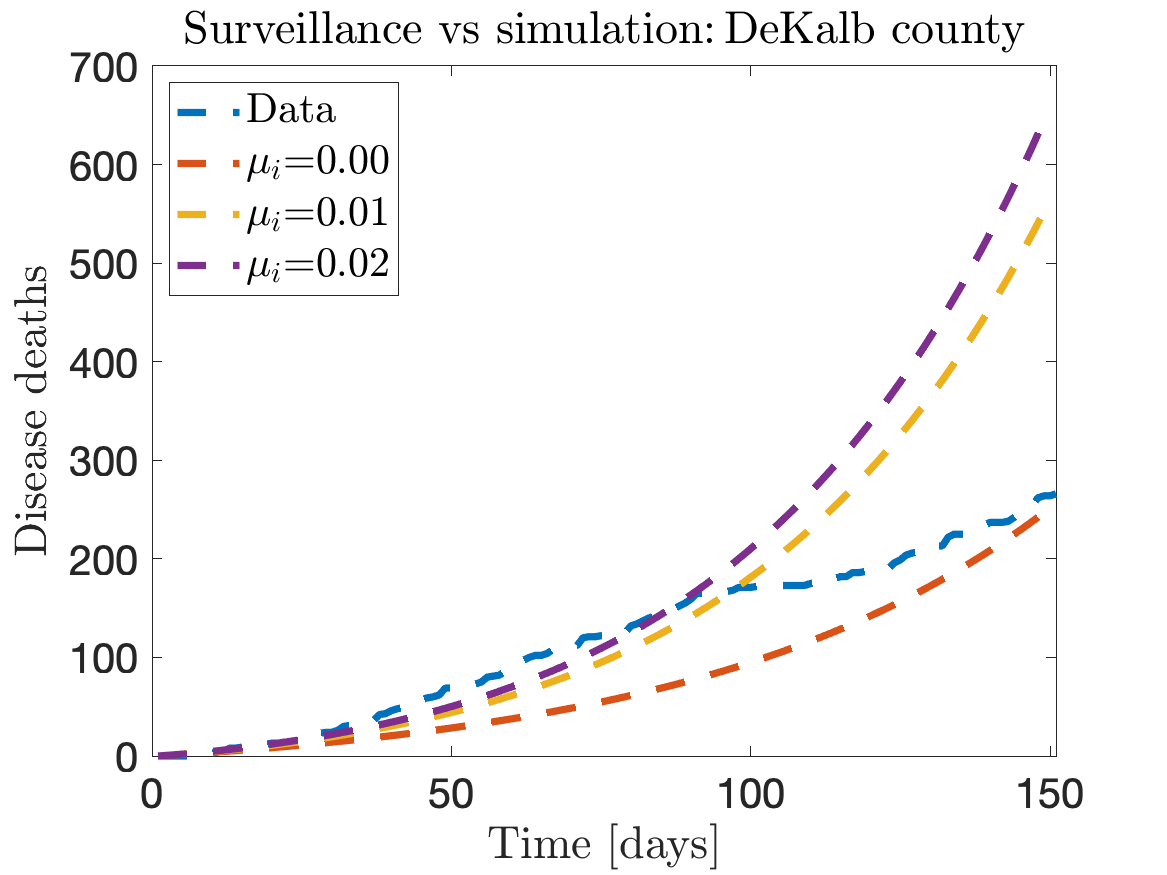}
  \caption{Georgia, DeKalb County}
  \label{fig:Dekalb}
\end{minipage}%
\begin{minipage}{.33\textwidth}
  \centering
  \includegraphics[width=.8\linewidth]{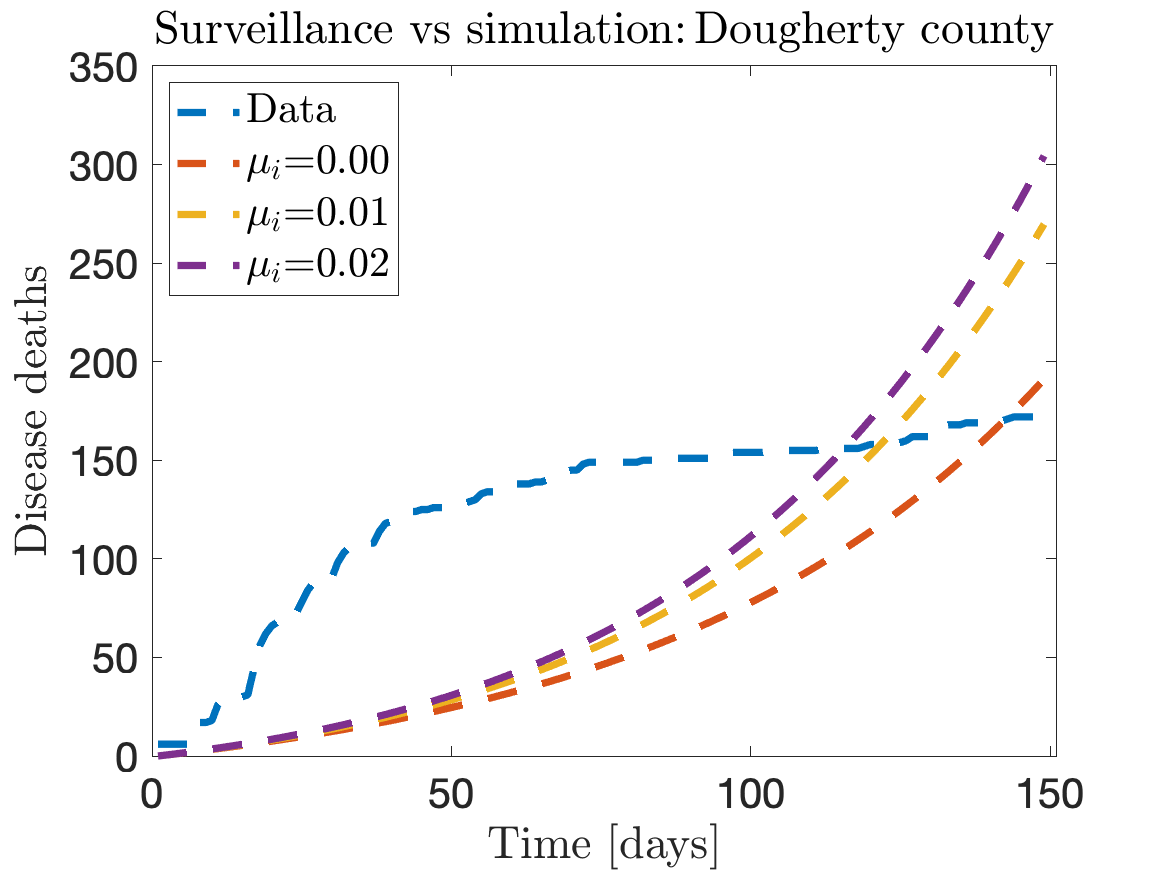}
\caption{Georgia, Dougherty County}
  \label{fig:Dougherty}
\end{minipage}
\begin{minipage}{.33\textwidth}
  \centering
  \includegraphics[width=.8\linewidth]{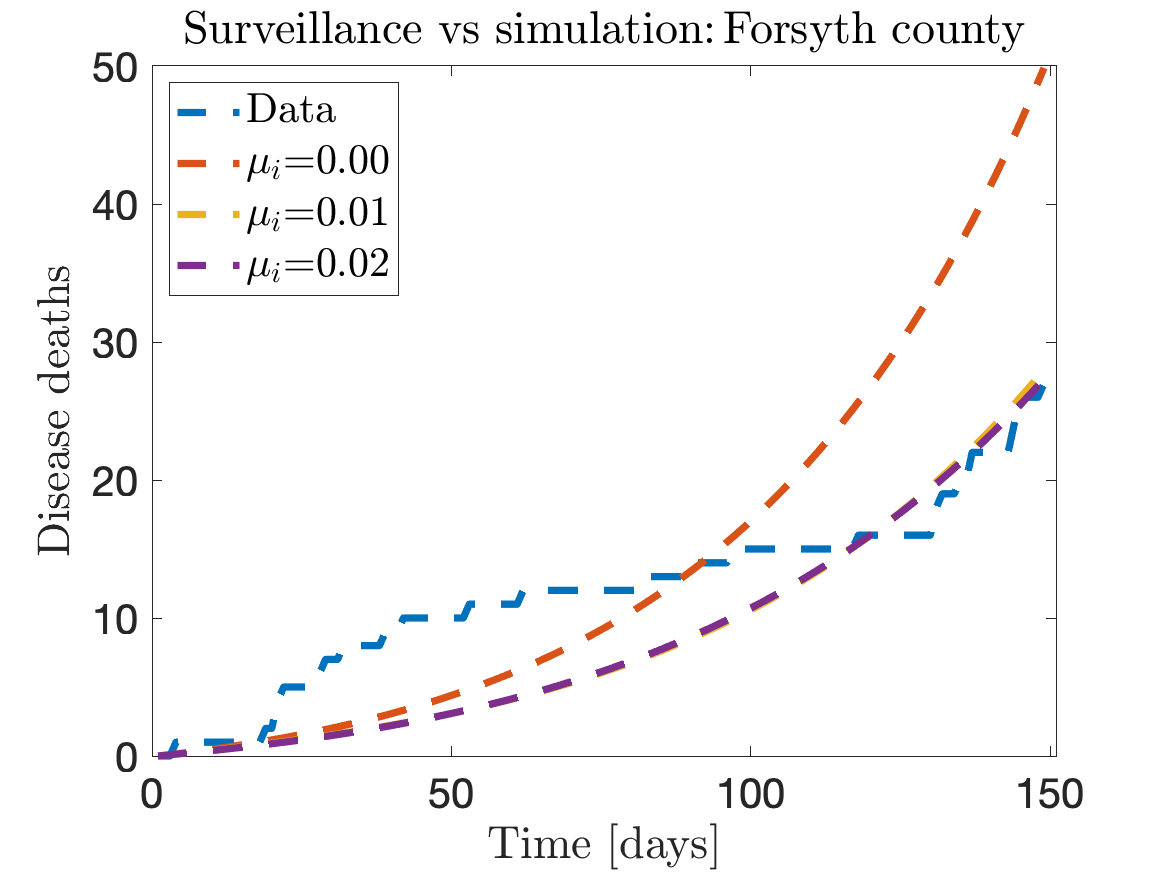}
\caption{Georgia, Forsyth County}
  \label{fig:Forsyth}
\end{minipage}
\end{figure}

\begin{figure}[ht!]
\centering
\begin{minipage}{.33\textwidth}
  \centering
  \includegraphics[width=.8\linewidth]{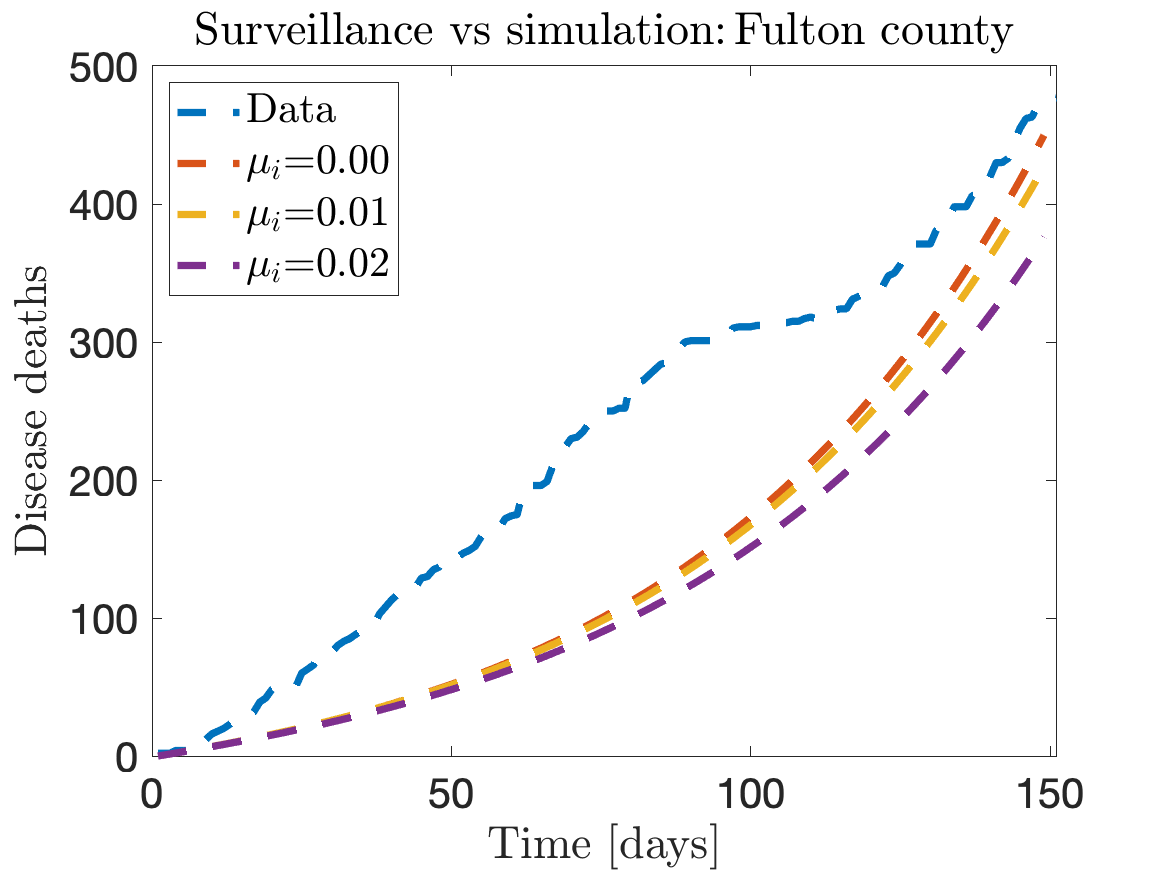}
  \caption{Georgia, Fulton County}
  \label{fig:Fulton}
\end{minipage}%
\begin{minipage}{.33\textwidth}
  \centering
  \includegraphics[width=.8\linewidth]{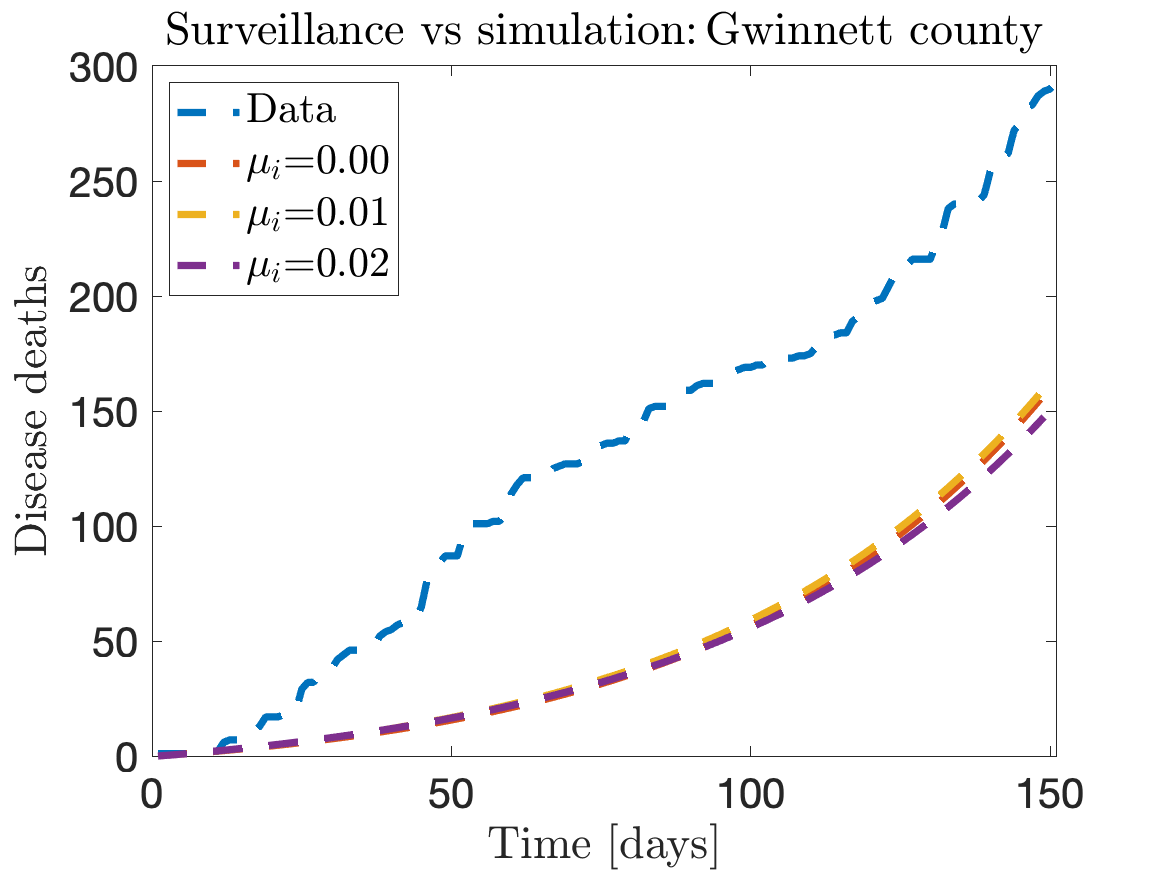}
\caption{Georgia, Gwinnett County}
  \label{fig:Gwinett}
\end{minipage}
\begin{minipage}{.33\textwidth}
  \centering
  \includegraphics[width=.8\linewidth]{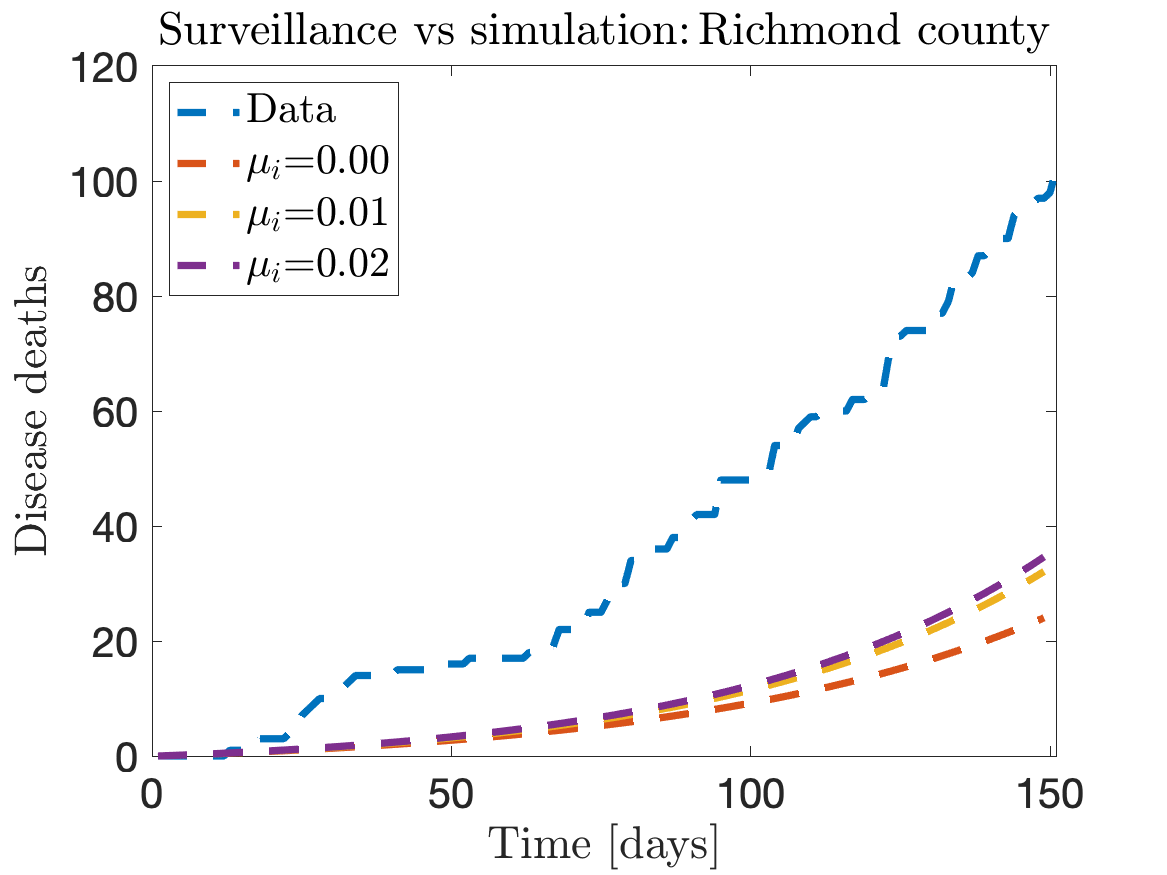}
\caption{Georgia, Richmond County}
  \label{fig:Richmond}
\end{minipage}
\end{figure}

We plot the results in Figs \ref{fig:Bibb}-\ref{fig:Richmond} for Bibb, Chatham, Cherokee, Clarke, Clayton, Cobb, DeKalb, Dougherty, Forsyth, Fulton, Gwinnett, and Richmond counties, respectively. For a qualitative comparison, we compared a fraction (10\%) of the removed compartment with the surveillance data for daily mortality. This comparison was made due to a lack of reliable testing data over the considered period to compare incidence data; this is consistent with simulations performed in \cite{viguerie2022coupled, grave2021assessing}, which also compared simulation outputs with surveillance data for daily mortality. However, as with Lombardy, the results should be interpreted with caution and primarily focused on comparing specific trends and behavior, and not on a purely quantitative level.  
\par Based on the results from \cite{grave2021assessing}, we are particularly interested in whether the chemotaxis model can increase transmission in Bibb, Chatham, and Richmond counties. The results suggest that this is indeed the case. For Bibb county (Fig. \ref{fig:Bibb}), the case of $\mu_i=0.02$ results in over double the transmission compared to the purely-diffusive case ($\mu_i=0.00$). Similar results were observed in Chatham (Fig. \ref{fig:Chatham}) and Richmond (Fig. \ref{fig:Richmond}) counties. Furthermore, the gap between the chemotaxis and non-chemotaxis cases appears to be increasing over time, suggesting that the chemotaxis model will outperform the non-chemotaxis model significantly over the longer term. 
\par We are also interested in the cases of Cherokee (Fig. \ref{fig:Cherokee}) and Forsyth (Fig. \ref{fig:Forsyth}) counties, in which the previously-employed purely diffusive models  \textit{overestimate} transmission \cite{grave2021assessing}. The chemotaxis model significantly reduces transmission in these cases, with the $\mu_i=0.02$ approximately halving the transmission observed in the purely diffusive case of $\mu_i=0.00$. 
\par The results in these specific cases suggest that the chemotaxis model can provide better results in specific instances where purely diffusive models failed to produce behavior consistent with reality. We now turn our attention to the other simulated counties to see the effect of chemotaxis when used in situations where purely diffusive models performed satisfactorily in \cite{grave2021assessing}. Chemotaxis results in significantly increased transmission in Clarke (Fig. \ref{fig:Clarke}), Cobb (Fig. \ref{fig:Cobb}), DeKalb (Fig. \ref{fig:Dekalb}), and Dougherty (Fig. \ref{fig:Dougherty})  counties. This transmission increase is unsurprising, as these areas are all population centers with less-populated areas in the immediate vicinity; therefore, we expect chemotaxis to have a significant effect due to the presence of population gradients. In contrast, in Clayton (Fig. \ref{fig:Clayton}), Fulton (Fig. \ref{fig:Fulton}), and Gwinnett (Fig. \ref{fig:Gwinett}) counties, chemotaxis had little effect. There was essentially no difference for Clayton and Gwinnett counties compared to the purely diffusive case; in Fulton county, chemotaxis resulted in slightly reduced transmission. This reduction is again unsurprising; while significant populations exist in each of these counties, they generally are surrounded by areas of similar population density, and therefore, the chemotaxis effect is small.

\begin{figure}[ht!]
    \centering
    \includegraphics[width=\textwidth]{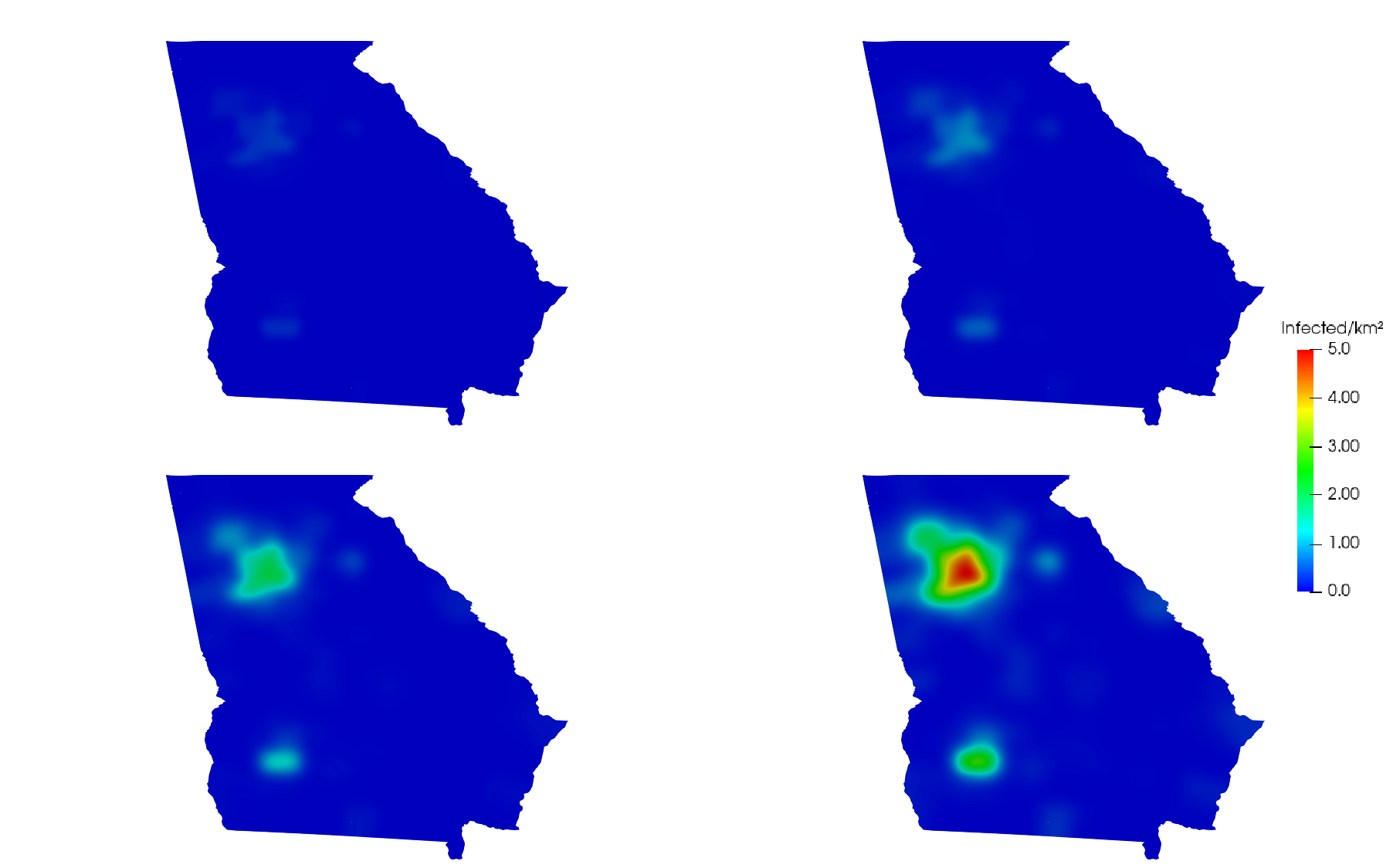}
    \caption{Days 50 (top-left), 100 (top-right), 150 (bottom-left) and 250 (bottom-right) of the Georgia simulation for $\mu_i=0.0$ and $\nu_i=0.25$ km$^2 \cdot$ Days $^{-1}$.} 
    \label{fig:georChem0Diff1}
\end{figure}

\begin{figure}[ht!]
    \centering
    \includegraphics[width=\textwidth]{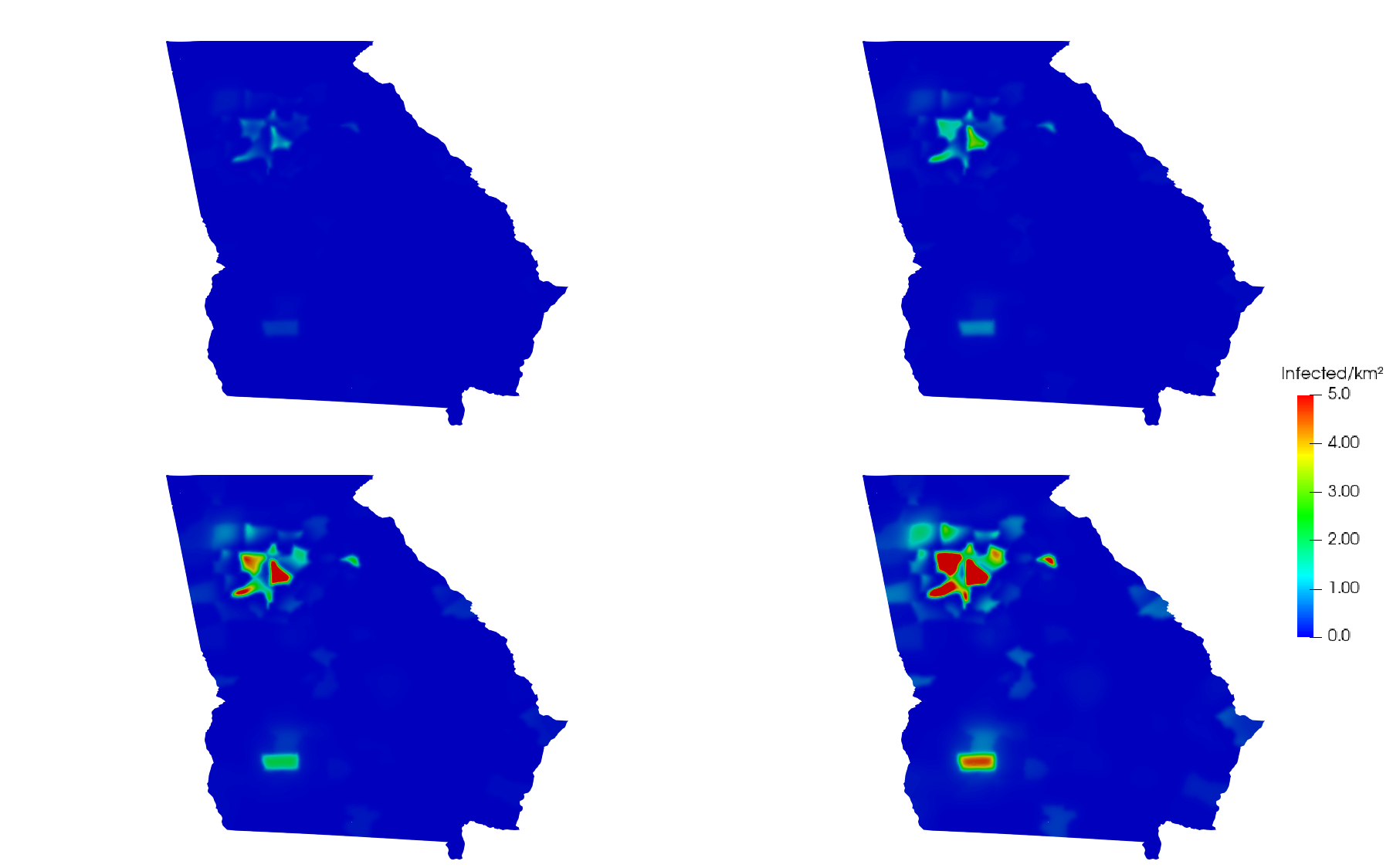}
    \caption{Days 50 (top-left), 100 (top-right), 150 (bottom-left) and 250 (bottom-right) of the Georgia simulation for $\mu_i=0.01$ and $\nu_i=0.25$ km$^2 \cdot$ Days $^{-1}$.} 
    \label{fig:georChem1Diff1}
\end{figure}

\begin{figure}[ht!]
    \centering
    \includegraphics[width=\textwidth]{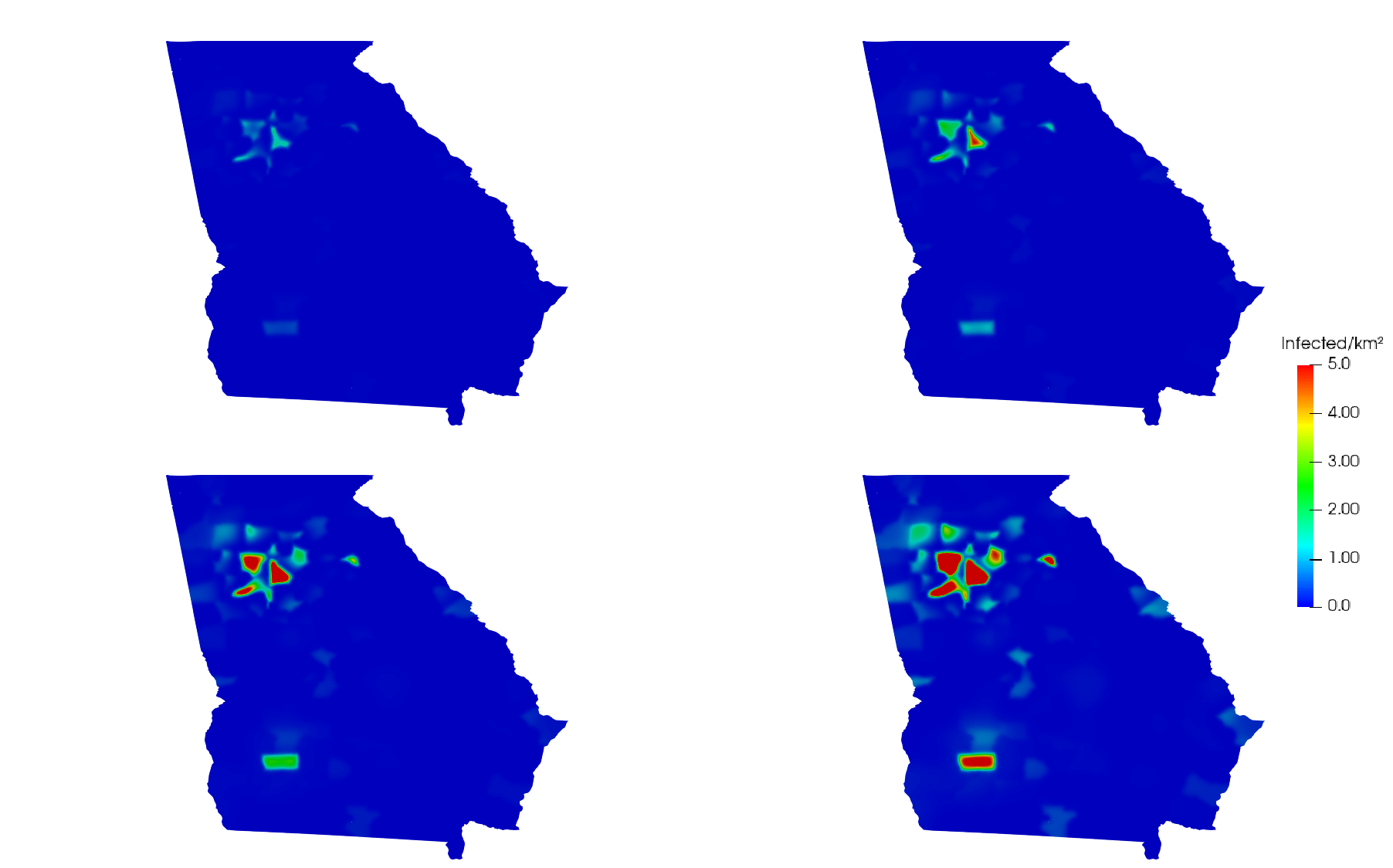}
    \caption{Days 50 (top-left), 100 (top-right), 150 (bottom-left) and 250 (bottom-right) of the Georgia simulation for $\mu_i=0.02$ and $\nu_i=0.25$ km$^2 \cdot$ Days $^{-1}$.} 
    \label{fig:georChem2Diff25}
\end{figure}

\begin{figure}[ht!]
    \centering
    \includegraphics[width=0.6\textwidth]{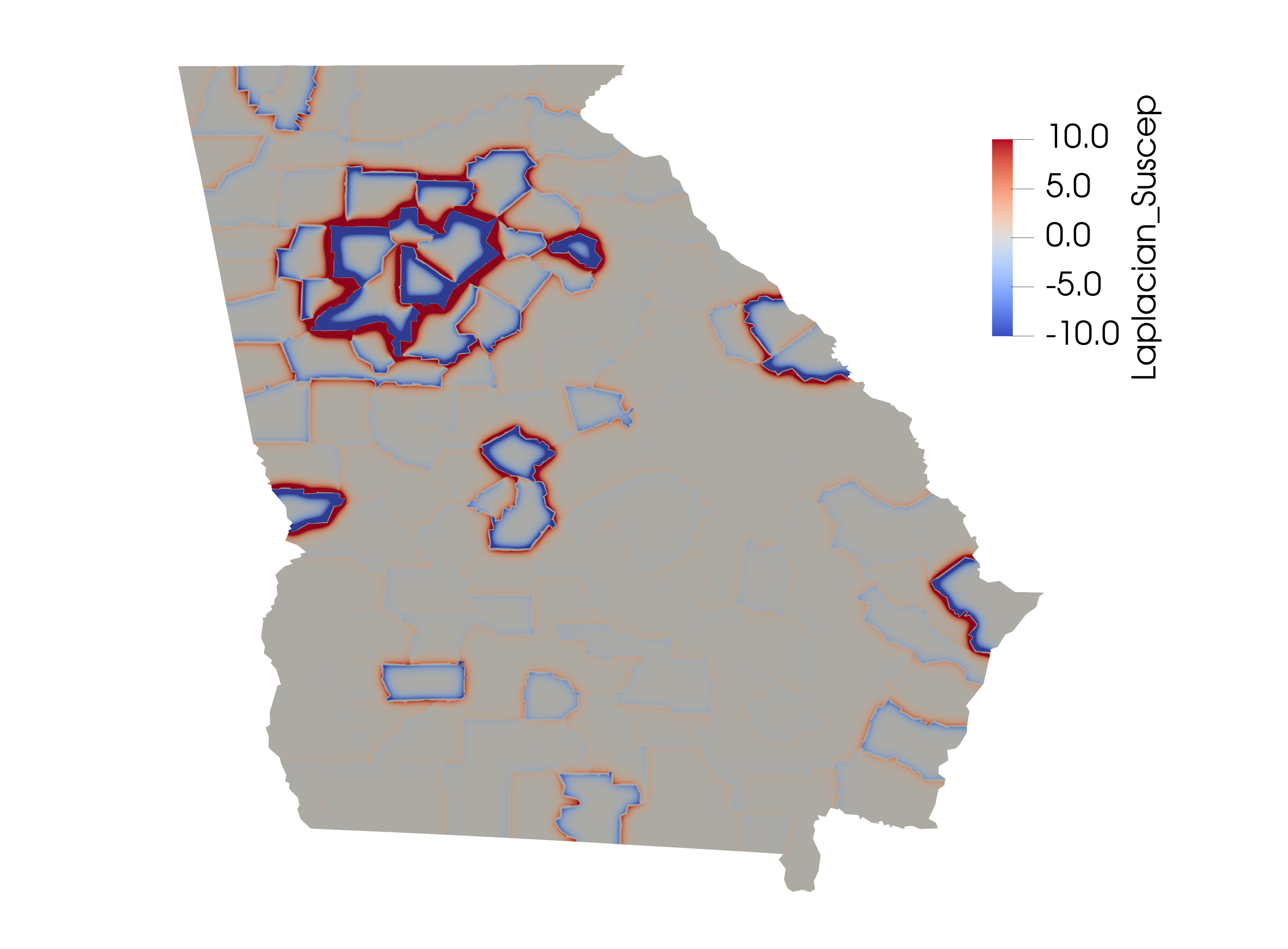}
    \caption{The Laplacian $\Delta s$ of the initial susceptible population $s$ in Georgia. Large negative values indicate areas particularly vulnerable to the epidemic. }
    \label{fig:georLaplace}
\end{figure}

\section{Conclusions and Future Directions}
This paper extends the study of a new PDE model for the spatiotemporal transmission of airborne diseases in human populations. We extended the mathematical analysis of the model, with an emphasis on results that connect the model's physical behavior to its mathematical structure in an interpretable, policy-actionable way. In particular, we derived spatially-aware analogues of the basic reproduction number, which establish that interplay between diffusive and chemotactic processes plays an important role in local transmission dynamics. Regions where the chemotaxis effect significantly exceeds diffusion can act as “hotspots” of infection, purely by virtue of spatial effects. These phenomena are independent of the standard transmission and recovery processes typically captured by reproduction numbers. Our findings underscore the potential impact of the Laplacian of the susceptible population density on disease transmission. Put simply, we find that infection hotspots are driven by local population density maxima, a potentially useful insight for public health practitioners.

\par We then conducted a series of numerical simulations in the Italian region of Lombardy and the U.S. state of Georgia. In addition to corroborating our mathematical results, these simulations suggest that a chemotaxis-based model may better capture airborne infectious disease spread in human populations than a purely diffusive model. Qualitative comparisons with COVID-19 surveillance data indicate that the chemotaxis model captures certain real-world dynamics more accurately, including the rapid spread of disease into population centers and the faster rates of transmission therein. Moreover, the chemotaxis model can produce spatially isolated “pockets” of infection in remote areas—a phenomenon observed in practice but not easily replicated by standard diffusive approaches without explicit reparameterization.

\par This work establishes further proof of concept for using chemotaxis-based models to describe airborne infectious disease transmission in human populations. There remain several important directions for further research. First, while our focus here was chiefly on qualitative comparisons with surveillance data, future studies should incorporate chemotaxis into models with more complex compartmental structures and refined parameter estimations to enable quantitative validation against epidemiological data. Second, introducing the chemotaxis parameter raises numerical challenges absent in the purely diffusive setting; to stabilize the simulations presented here, we employed low-order streamline-diffusion techniques designed for convection-dominated problems. Further efforts to develop order-preserving schemes or other stabilization strategies specifically tailored to chemotaxis-type PDEs would be valuable. One may look for instance, toward extending the results shown in \cite{kirk2009parallel}. Finally, although the two-dimensional, three-compartment model used here was computationally feasible, increasing the number of compartments can quickly inflate the dimensionality. In such situations, operator-splitting approaches—especially those implemented in parallel—may be necessary to keep simulations tractable. Future studies will explore the application of these methods to large-scale, chemotaxis-driven models. Finally, other dynamics, including nonlocal transmission \cite{grave2022modeling}, multiscale formulations \cite{albi2022kinetic}, hybrid models combining agent-based and compartmental formulations \cite{gopalappa2023new}, and other extensions, may be incorporated alongside the basic chemotactic structure studied herein. Finally, generalizing the local analysis shown here by relaxing the regularity conditions is also a worthwhile direction for future work.  

\bibliographystyle{unsrt}  
\bibliography{references}

\end{document}